\documentclass[journal]{IEEEtran}

\usepackage{acronym}
\usepackage{amsfonts}
\usepackage[dvips]{graphicx}
\usepackage{times}
\usepackage{cite}
\usepackage{amsmath}
\usepackage{array}
\usepackage{amssymb}
\usepackage{stfloats}
\usepackage{diagbox}
\usepackage{graphicx}
\usepackage{footnote}
\usepackage{amsthm}
\usepackage{booktabs}
\usepackage{array}
\usepackage[ruled,vlined]{algorithm2e}
\usepackage{subeqnarray}
\usepackage{cases}
\usepackage{threeparttable}
\usepackage{color}
\usepackage{epstopdf}
\usepackage{makecell}
\usepackage{multirow}
\usepackage{tabularx}
\usepackage{enumerate}
\usepackage{multicol}
\usepackage{subfig}
\usepackage{caption}

\def\bb0{{\mathbb{0}}}

\def\bb{{\mathbf{b}}}

\def\bm{{\mathbf{m}}}

\def\b0{{\mathbf{0}}}

\def\sf0{{\mathsf{0}}}

\def\rm0{{\mathrm{0}}}

\acrodef{CSI}[CSI]{channel state information}
\acrodef{CSIT}[CSIT]{channel state information at the transmitter}
\acrodef{CSIR}[CSIR]{channel state information at the receiver}
\acrodef{MIMO}[MIMO]{multiple-input multiple-output}
\acrodef{SISO}[SISO]{single-input single-output}
\acrodef{MISO}[MISO]{multiple-input single-output}
\acrodef{SIMO}[SIMO]{single-input multiple-output}
\acrodef{ADCs}[ADCs]{analog-to-digital convertors}
\acrodef{SNR}[SNR]{signal-to-noise ratio}
\acrodef{AWGN}[AWGN]{additive white Gaussian noise}
\acrodef{MRT}[MRT]{maximal ratio transmission}
\acrodef{DFT}[DFT]{Discrete Fourier Transform}
\acrodef{ULA}[ULA]{uniform linear array}
\acrodef{UPA}[UPA]{uniform planar array}
\acrodef{LS}[LS]{least squares}
\acrodef{ALMMSE}[ALMMSE]{approximate linear minimum mean squared error}
\acrodef{QIHT}[QIHT]{quantized iterative hard thresholding}
\acrodef{QIST}[QIST]{quantized iterative soft thresholding}
\acrodef{SVD}[SVD]{singular value decomposition}

\usepackage{url}
\usepackage{hyperref}
\usepackage{bm}
\usepackage{threeparttable}
\hypersetup{draft}

\usepackage{caption}
\ifCLASSINFOpdf

\else
\fi
\begin{document}

\title{Joint Channel Estimation and Signal Detection for MIMO-OFDM: A Novel Data-Aided Approach with Reduced Computational Overhead}
\author{Xinjie Li,~\IEEEmembership{Graduate Student Member,~IEEE,}
       Jing Zhang,~\IEEEmembership{Member,~IEEE,}
       Xingyu Zhou,~\IEEEmembership{Graduate Student Member,~IEEE,}
       Chao-Kai Wen,~\IEEEmembership{Fellow,~IEEE,}
       and Shi Jin,~\IEEEmembership{Fellow,~IEEE}
\thanks{X. Li, J. Zhang, X. Zhou, and S. Jin are with the National Mobile Communications Research Laboratory, Southeast University, Nanjing 210096, China
(e-mail: lixinjie@seu.edu.cn; jingzhang@seu.edu.cn; \protect \url{xy_zhou@seu.edu.cn}; jinshi@seu.edu.cn).}
\thanks{C.-K. Wen is with Institute of Communications Engineering, National Sun Yat-sen University, Kaohsiung 80424, Taiwan
(e-mail: chaokai.wen@mail.nsysu.edu.tw).}
}

\maketitle

\begin{abstract}
The acquisition of channel state information (CSI) is essential in MIMO-OFDM communication systems. Data-aided enhanced receivers, by incorporating domain knowledge, effectively mitigate performance degradation caused by imperfect CSI, particularly in dynamic wireless environments. However, existing methodologies face notable challenges: they either refine channel estimates within MIMO subsystems separately, which proves ineffective due to deviations from assumptions regarding the time-varying nature of channels, or fully exploit the time-frequency characteristics but incur significantly high computational overhead due to dimensional concatenation. To address these issues, this study introduces a novel data-aided method aimed at reducing complexity, particularly suited for fast-fading scenarios in fifth-generation (5G) and beyond networks. We derive a general form of a data-aided linear minimum mean-square error (LMMSE)-based algorithm, optimized for iterative joint channel estimation and signal detection. Additionally, we propose a computationally efficient alternative to this algorithm, which achieves comparable performance with significantly reduced complexity. Empirical evaluations reveal that our proposed algorithms outperform several state-of-the-art approaches across various MIMO-OFDM configurations, pilot sequence lengths, and in the presence of time variability. Comparative analysis with basis expansion model-based iterative receivers highlights the superiority of our algorithms in achieving an effective trade-off between accuracy and computational complexity.

\end{abstract}


\IEEEpeerreviewmaketitle

\section{Introduction}

Multiple-input-multiple-output orthogonal frequency division multiplexing (MIMO-OFDM), which combines spatial multiplexing and flat fading channels, has been a foundational technology since the advent of fourth-generation mobile cellular wireless systems. With the transition to fifth-generation (5G) and beyond, the New Radio (NR) architecture introduces a resource block (RB)-based frame structure that accommodates multiple numerologies for MIMO-OFDM transmissions \cite{trad5GNR1, trad5GNR2}.
 
Channel estimation and signal detection are critical components of MIMO-OFDM receivers. Channel frequency responses (CFRs) are typically estimated using training pilots, with a comb-type pattern facilitating the tracking of time-varying channel characteristics in realistic wireless environments \cite{CombSong, CombKhan}. Extrapolation or interpolation techniques \cite{tradOFDMCELiu}, particularly the linear minimum mean square error (LMMSE) method, are commonly employed to estimate the CFRs of data symbols. For signal detection, message-passing-based detectors, such as approximate message passing (AMP) and expectation propagation (EP) \cite{tradMIMOSDJeo, tradMIMOSDWu, tradMIMOSDCes}, provide a favorable balance between performance and complexity, leading to their widespread adoption.

However, detection performance degrades significantly in the presence of imperfect channel state information (CSI) \cite{tradMIMOSDGha}. In recent years, numerous studies have sought to enhance traditional receiver designs using deep learning (DL). Specifically, data-driven approaches \cite{dlDDYe, dlDDHon, dlDDXin} leverage neural networks to recover transmitted bits directly from the received signal, bypassing the need for domain knowledge. While these neural network-based methods improve receiver performance through joint optimization, they come with substantial computational costs. Moreover, pretrained neural networks often experience significant performance degradation in real-world dynamic channels \cite{dlDDFu, dlDDShang, dlDDJu}. These challenges in complexity and generalization underscore the need for improved designs that integrate domain knowledge, spurring a resurgence in the {\bf J}oint {\bf C}hannel estimation and signal {\bf D}etection (JCD) framework.
 
The JCD framework, applicable to a wide range of communication systems, offers a promising solution to mitigate performance degradation caused by imperfect CSI. Within this framework, data-aided channel estimation leverages \emph{a priori} information from data estimates to enhance accuracy, thereby improving signal detection reliability. Typically, the decoder is integrated into the JCD iterations for error correction, with the output extrinsic information enabling more precise data feedback. The integration of JCD within receiver designs has been extensively investigated in \cite{daJOINTari,daCOFDMGeng, daCMIMOOFDMYli, daOTFSHuang, daNLZhang}.

Despite significant progress, recent works emphasize variants of the joint optimization problem to further advance JCD designs. Some approaches construct prior assumptions about channel characteristics to approximate the posterior distribution \cite{daMIMOKar, daMIMONgu, daMIMOWang, daMIMOWei}. For example, in \cite{daMIMOKar, daMIMONgu, daMIMOWang}, the joint posterior distribution of channels and data symbols is factorized, and bilinear Bayesian inference is applied using Gaussian \cite{daMIMOKar, daMIMONgu} or Bernoulli-Gaussian \cite{daMIMOWang} priors. Similarly, \cite{daMIMOWei} approximates the posterior via a Gaussian mixture model (GMM), introducing subspace projection methods to enhance the GMM-based data-aided estimator. Other variants enhance JCD receivers by incorporating model-driven DL methods \cite{daMIMOHe, daMIMOZhang, daMIMOZhangE, daMIMOSun, daMIMOBha, daMIMOYang}. Specifically, \cite{daMIMOHe, daMIMOZhang} develop LMMSE-based data-aided channel estimation and refine message-passing signal detection with trainable hyperparameters to improve convergence and performance. Deep unfolding techniques are integrated in \cite{daMIMOZhangE, daMIMOSun, daMIMOBha} using algorithms such as generalized expectation maximization \cite{daMIMOZhangE} and the alternating direction method of multipliers \cite{daMIMOSun, daMIMOBha}, achieving improved detection performance with data-aided estimation derived via variational inference. \cite{daMIMOYang} integrates neural network-enhanced designs within bilinear Gaussian belief propagation iterations to perform channel extrapolation and refine prior knowledge for data-aided estimation.

While these JCD designs \cite{daMIMOKar, daMIMONgu, daMIMOWang, daMIMOWei, daMIMOHe, daMIMOZhang, daMIMOZhangE, daMIMOSun, daMIMOBha, daMIMOYang} provide valuable insights, they are predominantly developed under block fading channel assumptions, making them unsuitable for high-mobility scenarios in 5G and beyond networks \cite{trad5GNR1, trad5GNR3}, where quasi-static models fail. In contrast, the comb-type pilot pattern effectively captures the symbol-by-symbol time-varying nature of wireless channels. However, existing data-aided estimators face significant complexity challenges.
 
For instance, works such as \cite{daMIMOOFDMNov, daMIMOOFDMWu, daMIMOOFDMKir} address scalability by constructing equivalent systems that concatenate information across time, frequency, and spatial domains, approximating the probability density function (pdf) using message-passing techniques. This concatenation, however, incurs high computational costs. Complexity reduction is achieved in \cite{daCOMPKas} by employing the basis expansion model (BEM), which uses discrete prolate spheroidal (DPS) sequences \cite{daCOMPHam} and B-spline basis functions \cite{daCOMPZak, daCOMPKha} to model fading characteristics. Nevertheless, time-frequency concatenation in data-aided estimation remains computationally expensive, and reliance on decision criteria for true data estimation errors introduces potential inaccuracies. Thus, further investigation is needed into data-aided channel estimation techniques that support subcarrier-specific allocation of pilots and symbols in MIMO-OFDM systems. This exploration is crucial for developing practical schemes that balance flexibility, performance, and complexity reduction in time-selective fading environments.

In this paper, we propose a JCD framework for MIMO-OFDM receivers that supports flexible deployment of various detectors and optionally incorporates a decoding module. The EP detector is applied, and the decoding module is activated only after convergence in the JCD process to minimize computational load. Unlike previous approaches reliant on block-fading assumptions \cite{daMIMOKar, daMIMONgu, daMIMOWang, daMIMOWei, daMIMOHe, daMIMOZhang, daMIMOZhangE, daMIMOSun, daMIMOBha, daMIMOYang} or incurring substantial computational overhead \cite{daMIMOOFDMNov, daMIMOOFDMWu, daMIMOOFDMKir, daCOMPKas}, the proposed data-aided LMMSE channel estimation method is applicable to various time-selective fading scenarios. A closed-form solution is derived by incorporating true detection error statistics, and an equivalent representation is proposed to significantly reduce complexity. The main contributions of this paper are summarized as follows: 
\begin{itemize}
\item The general form of data-aided LMMSE-based channel estimation within the JCD structure, termed \emph{MJCD-LMMSE}, is derived. \emph{MJCD-LMMSE} leverages the estimated data symbols provided by the EP detector in the previous JCD layer to compute refined channel estimations, thereby providing more accurate coefficients for subsequent signal detection. Extrinsic log-likelihood ratios (LLRs), based on EP estimates from the final JCD iteration, are computed for channel decoding.

\item To reduce the complexity of implementation in standard MIMO-OFDM configurations, a novel equivalent representation, termed \emph{OJCD-LMMSE}, is introduced. \emph{OJCD-LMMSE} decouples spatial-frequency correlation properties into individual OFDM subsystems, enabling separate quantification of interference statistics. This approach results in an alternative solution that significantly reduces complexity compared to \emph{MJCD-LMMSE}.

\item Numerical simulations demonstrate that the \emph{MJCD-LMMSE}-based approach significantly outperforms traditional methods. Moreover, substituting \emph{MJCD-LMMSE} with \emph{OJCD-LMMSE} in the proposed framework delivers similar performance with remarkably lower computational cost. Evaluations under various time-varying conditions confirm the practical effectiveness of the proposed design in realistic wireless transmissions, highlighting its conspicuous advantage in balancing performance and complexity.

\end{itemize}

The rest of this paper is organized as follows. MIMO-OFDM system model including traditional LMMSE estimation and EP detection based receiver is introduced in Section \ref{sec_system_model}. In Section \ref{sec_MIMO_OFDM_CE}, \emph{MJCD-LMMSE} is derived for data-aided channel estimation in our proposed JCD structure, and the equivalent low-complexity algorithm \emph{OJCD-LMMSE} is proposed in Section \ref{sec_OFDM_CE}. Numerical results are represented in Section \ref{sec_simulation}, and Section \ref{sec_conclusion} finally concludes the paper.

\emph{Notations:} Superscripts ${(  \cdot  )^T}$ and ${(  \cdot  )^{\mathrm{H}}}$ denote the transpose and conjugate transpose respectively. $z^*$ and $| z |$ denote the complex conjugate and modulus of a complex number $z$. The expectation operator is denoted by $\mathbb{E}\{  \cdot  \}$, while $\text{Var} \{  \cdot  \}$ indicates the variance. $\mathbf{I}$ is the identity matrix, $\mathbf{0}$ represents the zero matrix, and $\mathcal{N}( {{\mathbf{0}},\sigma _w^2{\mathbf{I}}} )$ indicates Gaussian random variables with zero mean and variance $\sigma _w^2$. $\text{diag} ( \mathbf{x} )$ returns a diagonal matrix with $\mathbf{x}$ on the main diagonal. In addition, $\otimes$ and $\odot$ represent the Kronecker product and Hadamard product respectively, and $\lfloor {\cdot} \rfloor$ represents the floor operator.

\section{System Model}
\label{sec_system_model}
\par
\begin{figure*}[htbp!]
  \centerline{\includegraphics[width=1.0\textwidth]{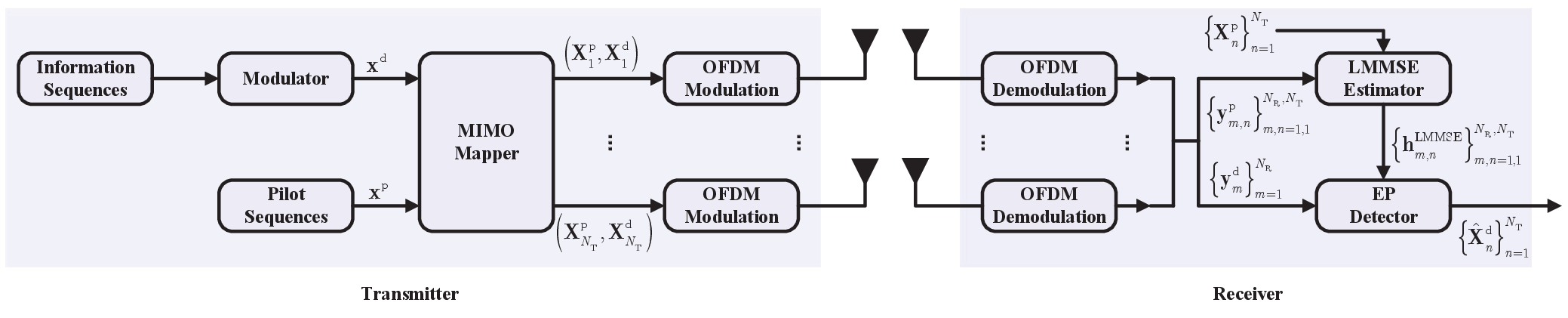}}
  \caption{~Block diagram of MIMO-OFDM receiver with traditional LMMSE channel estimation and EP signal detection.}
  \label{fig_system_model}
  \vspace{-1em}
\end{figure*}

A MIMO-OFDM system configured with $N_{\text{R}} \times N_{\text{T}}$ antennas and $K$ subcarriers is considered, where traditional LMMSE interpolator is adopted for comb-type pilot-assisted channel estimation and EP detector is utilized in signal detection, as illustrated in Fig. \ref{fig_system_model}.

\subsection{MIMO-OFDM}

An OFDM frame consisting of a pilot block and a data block is formed and transmitted at each antenna, under the duration of which the corresponding channel coefficients are assumed to be constant. $P$ out of $K$ subcarriers are chosen for inserting pilot sequences, while other subcarriers are used for data transmission. For more accurate estimation, the subcarriers taken up by pilot symbols are spaced evenly in $\lfloor {\frac{K}{P}} \rfloor$ intervals. Moreover, pilot sequences transmitted at different antennas are required to be orthogonal. The received signals corresponding to ${N_{\text{R}}}{N_{\text{T}}}$ sets of OFDM channels are demanded.

Specifically, the output at the $m$-th receiving antenna corresponding to the $n$-th transmitted pilot block ${\mathbf{y}}_{m,n}^{\text{p}} = {[ {y_{m,n}^{{p_1}}, \ldots ,y_{m,n}^{{p_P}}} ]^T} \in {\mathbb{C}^{P \times 1}}$ can be expressed as
\begin{equation}
  {\mathbf{y}}_{m,n}^{\text{p}} = {\mathbf{X}}_n^{\text{p}}{\mathbf{h}}_{m,n}^{\text{p}} + {\mathbf{w}}_{m,n}^{\text{p}}
  \label{equ_y_mn_p}
\end{equation}
for $m=1,2,\dots, N_{\text{R}}$ and $n=1,2,\dots, N_{\text{T}}$, where $\{ {{p_k}} \}_{k = 1}^P$ denotes the index for pilot subcarriers. The transmitted pilot block is denoted as ${\mathbf{X}}_n^{\text{p}} = {\text{diag}} ( {{\mathbf{x}}_n^{\text{p}}}  ) = {\text{diag}} ( {x_n^{{p_1}}, \ldots ,x_n^{{p_P}}}  ) \in {\mathbb{C}^{P \times P}}$, and the corresponding channel coefficients are ${\mathbf{h}}_{m,n}^{\text{p}} = {[ {h_{m,n}^{{p_1}}, \ldots ,h_{m,n}^{{p_P}}} ]^T}$. ${\mathbf{w}}_{m,n}^{\text{p}} = {[ {w_{m,n}^{{p_1}}, \ldots,w_{m,n}^{{p_P}}} ]^T}$ is an additive white Gaussian noise (AWGN) vector, i.e., ${\mathbf{w}}_{m,n}^{\text{p}} \sim \mathcal{N}( {{\mathbf{0}},\sigma _w^2{\mathbf{I}}} )$.
\par
As for data transmission, the information sequences are modulated with a complex $M$-ary quadrature amplitude modulation (QAM) constellation $\mathcal{A}$ and assembled as data blocks, which are represented as ${\mathbf{X}}_n^{\text{d}} = {\text{diag}}( {{\mathbf{x}}_n^{\text{d}}} ) = {\text{diag}}( {x_n^{{d_1}},\ldots,x_n^{{d_{K - P}}}} ) \in {\mathbb{C}^{( {K - P} ) \times ( {K - P} )}}$ for $n=1,2,\dots, N_{\text{T}}$. The data blocks are then transmitted over the channel simultaneously. The received data block at the $m$-th receiving antenna ${\mathbf{y}}_m^{\text{d}} = {[ {y_m^{{d_1}}, \ldots ,y_m^{{d_{K - P}}}} ]^T} \in {\mathbb{C}^{( {K - P} ) \times 1}}$ is denoted as
\begin{equation}
  {\mathbf{y}}_m^{\text{d}} = \sum\limits_{n = 1}^{{N_{\text{T}}}} {{\mathbf{X}}_n^{\text{d}}{\mathbf{h}}_{m,n}^{\text{d}}}  + {\mathbf{w}}_m^{\text{d}}
  \label{equ_y_m_d}
\end{equation}
for $m=1,2,\dots, N_{\text{R}}$. Similarly, $\{ {{d_k}} \}_{k = 1}^{K-P}$ denotes the index for data subcarriers, and the noise vector has independent components with zero-mean and $\sigma _w^2$-variance. The corresponding channel coefficients are ${\mathbf{h}}_{m,n}^{\text{d}} = {[ {h_{m,n}^{{d_1}}, \ldots ,h_{m,n}^{{d_{K - P}}}} ]^T}$. Notably, for specific ${d_k}$, the corresponding vector ${{\mathbf{y}}^{{d_k}}} = {[ {y_1^{{d_k}}, \ldots ,y_{{N_{\text{R}}}}^{{d_k}}} ]^T} \in {\mathbb{C}^{{N_{\text{R}}} \times 1}}$ consisting of ${N_{\text{R}}}$ components from different receiving antennas can also be represented as
\begin{equation}
  {{\mathbf{y}}^{{d_k}}} = {{\mathbf{H}}^{{d_k}}}{{\mathbf{x}}^{{d_k}}} + {{\mathbf{w}}^{{d_k}}},
  \label{equ_y_dk}
\end{equation}
where ${{\mathbf{x}}^{{d_k}}} = {[ {x_1^{{d_k}}, \ldots ,x_{{N_{\text{T}}}}^{{d_k}}} ]^T} \in {\mathbb{C}^{{N_{\text{T}}} \times 1}}$ and ${{\mathbf{H}}^{{d_k}}} \in {\mathbb{C}^{{N_{\text{R}}} \times {N_\text{T}}}}$. Consequently, at any specific data subcarrier, the receiving representation can be treated as MIMO receiving signal equivalently. The equivalent MIMO system can be represented as
\begin{equation}
  {\mathbf{y}} = {\mathbf{Hx}} + {\mathbf{w}},
  \label{equ_y_mimo}
\end{equation}
where the subcarrier index ${d_k}$ is omitted for simplicity.

\subsection{LMMSE-based Channel Estimation}
\par
Since only $P$ subcarriers are used for pilot transmission, channel estimation is expected to recover unknown channel coefficients at data subcarriers using acquirable information at pilot subcarriers. Based on least squares (LS) estimation at pilot subcarriers, LMMSE estimation can realize such a goal utilizing the frequency correlation in MIMO-OFDM channels. Specifically, the initial LS estimation at pilot subcarriers is
\begin{equation}
  {\mathbf{h}}_{m,n}^{{\text{LS}}} = {\left( {{\mathbf{X}}_n^{\text{p}}} \right)^{ - 1}}{\mathbf{y}}_{m,n}^{\text{p}},
  \label{equ_h_mn_LS}
\end{equation}
and the estimated channel coefficients at all subcarriers using LMMSE algorithm is
\begin{align}
  {\mathbf{h}}_{m,n}^{{\text{LMMSE}}} &= {{\mathbf{W}}_{{\text{LMMSE}}}}{\mathbf{h}}_{m,n}^{{\text{LS}}},
  \label{equ_h_mn_LMMSE} \\
  {{\mathbf{W}}_{{\text{LMMSE}}}} &= {{\mathbf{R}}_{{\mathbf{h}}{{\mathbf{h}}^{\text{p}}}}}{\left( {{{\mathbf{R}}_{{{\mathbf{h}}^{\text{p}}}{{\mathbf{h}}^{\text{p}}}}} + \sigma _w^2{\mathbf{I}}} \right)^{ - 1}},
  \label{equ_W_LMMSE}
\end{align}
where the correlation among subcarriers ${{\mathbf{R}}_{{\mathbf{h}}{{\mathbf{h}}^{\text{p}}}}} \in {\mathbb{C}^{K \times P}}$ and ${{\mathbf{R}}_{{{\mathbf{h}}^{\text{p}}}{{\mathbf{h}}^{\text{p}}}}}  \in {\mathbb{C}^{P \times P}}$ can be acquired from the channel correlation matrix in the frequency domain, i.e., $\mathbf{R}^{\text{Freq}}$.

\subsection{EP-based Signal Detection}
\par
After acquiring LMMSE estimation of ${N_{\text{R}}}{N_{\text{T}}}$ sets of OFDM channels, channel coefficients at data subcarriers are utilized during signal detection. According to (\ref{equ_y_dk}), at any data subcarrier ${d_k}$, the ${N_{\text{R}}}{N_{\text{T}}}$ estimated components can be reassembled as a matrix such that ${\hat{\mathbf{H}}^{{d_k}}} \in {\mathbb{C}^{{N_{\text{R}}} \times {N_T}}}$, which can be further simplified as ${\hat{\mathbf{H}}}$ during the operation of signal detection.
\par
EP approximates the posterior distribution with factorized Gaussian distributions as follows \cite{tradMIMOSDCes}
\begin{align}
  p\left( {{\mathbf{x}}|{\mathbf{y}}} \right) &\propto \mathcal{N}\left( {{\mathbf{y}};{\mathbf{Hx}},\sigma _w^2{{\mathbf{I}}_{{N_{\text{R}}}}}} \right) \cdot {\prod\limits_{n = 1}^{{N_{\text{T}}}} {{p_{\text{a}}}\left( {{x_n}} \right)} },
  \label{equ_p_x_y} \\
  q\left( {{\mathbf{x}}|\bm{\gamma} ,\bm{\Lambda} } \right) &\propto \mathcal{N}\left( {{\mathbf{y}};{\mathbf{Hx}},\sigma _w^2{{\mathbf{I}}_{{N_{\text{R}}}}}} \right) \cdot \mathcal{N}\left( {{\mathbf{x}};{\bm{\Lambda} ^{ - 1}}\bm{\gamma} ,{\bm{\Lambda}}} \right),
  \label{equ_q_x_gamma_lamta}
\end{align}
where ${p_{\text{a}}}( {{x_n}} )$ is the \emph{a priori} pdf of $\mathbf{x}$, and the EP solution (\ref{equ_q_x_gamma_lamta}) approximates (\ref{equ_p_x_y}) by recursively updating $( \bm{\gamma}, \bm{\Lambda} )$. After $T$ iterations, data estimates $\hat{\mathbf{x}}$ are output.

\begin{figure}[t]
  \begin{minipage}{3.5in}
    \centerline{\includegraphics[width=3.5in]{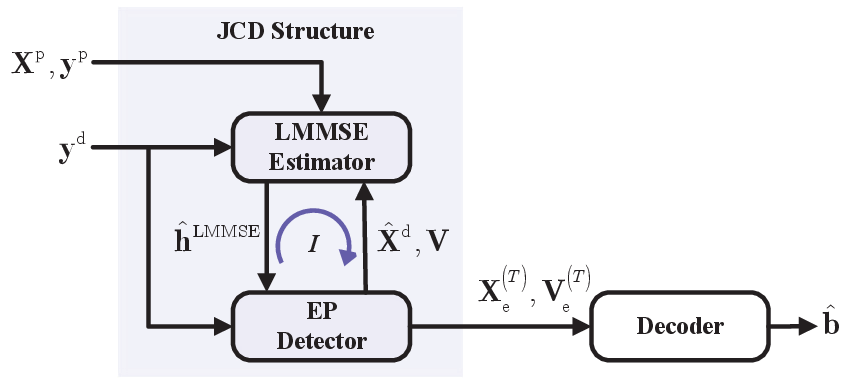}}
  \end{minipage}
  \caption{~Illustration of MIMO-OFDM receiver with the proposed JCD structure and channel decoder.}
  \label{fig_jcd_structure}
  \vspace{-1em}
\end{figure}

\begin{algorithm}[t]
  \SetAlgoLined
  \caption{EP Detector using imperfect CSI}
  \label{algo_ep}
  \KwIn{$\hat{\mathbf{H}}$, $\mathbf{y}$, $\sigma_w^2$, $T$, $\beta$}
  \KwOut{${\mathbf{x}}_{\text{p}}^{\left( T \right)}$}
  \SetKw{KwIni}{Initialize:}
  \KwIni $\gamma _i^{\left( 0 \right)} = 0$, $\lambda _i^{\left( 0 \right)} = E_{\text{s}}^{ - 1}$
  \BlankLine
	\For{$t = 1, \ldots ,T$}{
    Compute covariance and mean of the unnormalized Gaussian distribution:
		\begin{equation}
      \begin{aligned}
        \bm{\Sigma} ^{\left( t \right)} &= {\left( {\sigma _w^{ - 2}{{\hat{\mathbf{H}}}^T}{\hat{\mathbf{H}}} + {\bm{\lambda} ^{\left( {t - 1} \right)}}} \right)^{ - 1}}, \\
        \bm{\mu} ^{\left( t \right)} &= {\bm{\Sigma} ^{\left( t \right)}}\left( {\sigma _w^{ - 2}{{\hat{\mathbf{H}}}^T}{\mathbf{y}} + {\bm{\gamma} ^{\left( {t - 1} \right)}}} \right);
      \end{aligned}
      \label{ep_sigma_miu}
    \end{equation} \\
    Compute extrinsic covariance and mean of the cavity marginal:
    \begin{equation}
      \begin{aligned}
        {\mathbf{V}}_{\text{e}}^{\left( t \right)} &= \frac{{{\bm{\Sigma} ^{\left( t \right)}}}}{{1 - {\bm{\Sigma} ^{\left( t \right)}}{\bm{\lambda} ^{\left( {t - 1} \right)}}}}, \\
        {\mathbf{x}}_{\text{e}}^{\left( t \right)} &= {\mathbf{V}}_{\text{e}}^{\left( t \right)}\left( {\frac{{{\bm{\mu} ^{\left( t \right)}}}}{{{\bm{\Sigma} ^{\left( t \right)}}}} - {\bm{\gamma} ^{\left( {t - 1} \right)}}} \right);
      \end{aligned}
      \label{ep_Ve_xe}
    \end{equation} \\
    Compute the posterior mean and covariance:
    \begin{equation}
      \begin{aligned}
        {\mathbf{x}}_{\text{p}}^{\left( t \right)} &= \mathbb{E} \left\{ {{\mathbf{x}}|{\mathbf{x}}_{\text{e}}^{\left( t \right)},{\mathbf{V}}_{\text{e}}^{\left( t \right)}} \right\}, \\
        {\mathbf{V}}_{\text{p}}^{\left( t \right)} &= {\text{Var}}\left\{ {{\mathbf{x}}|{\mathbf{x}}_{\text{e}}^{\left( t \right)},{\mathbf{V}}_{\text{e}}^{\left( t \right)}} \right\};
      \label{ep_Vp_xp}
      \end{aligned}
    \end{equation} \\
    Refine the parameter pairs:
    \begin{equation}
      \begin{aligned}
        {\bm{\lambda} ^{\left( t \right)}} &= {\left( {{\mathbf{V}}_{\text{p}}^{\left( t \right)}} \right)^{ - 1}} - {\left( {{\mathbf{V}}_{\text{e}}^{\left( t \right)}} \right)^{ - 1}}, \\
      {\bm{\gamma} ^{\left( t \right)}} &= {\left( {\mathbf{V}_{\text{p}}^{\left( t \right)}} \right)^{ - 1}}{\mathbf{x}}_{\text{p}}^{\left( t \right)} - {\left( {{\mathbf{V}}_{\text{e}}^{\left( t \right)}} \right)^{ - 1}}{\mathbf{x}}_{\text{e}}^{\left( t \right)};
      \end{aligned}
      \label{ep_lamta_gamma}
    \end{equation}
		\If {$\lambda _i^{\left( t \right)} < 0$}
      {$\lambda _i^{\left( t \right)} = \lambda _i^{\left( t-1 \right)}$, $\gamma _i^{\left( t \right)} = \gamma _i^{\left( t-1 \right)}$;}
    Smooth parameter updates:
    \begin{equation}
      \begin{aligned}
        {\bm{\lambda} ^{\left( t \right)}} &= \beta {\bm{\lambda} ^{\left( t \right)}} + \left( {1 - \beta } \right){\bm{\lambda} ^{\left( {t - 1} \right)}}, \\
      {\bm{\gamma} ^{\left( t \right)}} &= \beta {\bm{\gamma} ^{\left( t \right)}} + \left( {1 - \beta } \right){\bm{\gamma} ^{\left( {t - 1} \right)}}.
      \end{aligned}
      \label{ep_damping}
    \end{equation}
	}
\end{algorithm}

\par
When the actual channel coefficients are not perfectly known, EP detection can be performed according to Algorithm \ref{algo_ep}. However, under such circumstances, degradation in detection performance is exhibited \cite{tradMIMOSDGha}. This performance degradation necessitates an improved design that takes the imperfect CSI into account. Specifically, the data-aided method is considered, using the estimated symbols for a more accurate estimation of channel coefficients, which in turn mitigates the influences induced by channel estimation error. The detailed design is illustrated in the next section.

\section{Proposed Channel Estimation}
\label{sec_MIMO_OFDM_CE}
\par
In this section, we propose the LMMSE-based data-aided method in MIMO-OFDM receiver, where JCD structure with $I$ iterations is considered, as shown in Fig. \ref{fig_jcd_structure}. The proposed data-aided channel estimation algorithm, combined with EP detector, is utilized in JCD-2 to JCD-$I$. Note that the combination of the methods in Section II-A and II-B, i.e. traditional LMMSE and EP, corresponds to the case that $I=1$. Moreover, channel decoding is performed only once to compute LLRs according to extrinsic information $( {{\mathbf{x}}_{\text{e}}^{( T )},{\mathbf{V}}_{\text{e}}^{( T )}} )$ in (\ref{ep_Ve_xe}) offered by EP detector at JCD-$I$. The proposed method, simplified as \emph{MJCD-LMMSE}, is presented as follows.

\subsection{MJCD-LMMSE}
\par
For the derivation of the estimation algorithm, we begin with the equivalent representation of MIMO-OFDM system. Specifically, consider MIMO-OFDM system as a ``MIMO'' system with large scale, where data blocks from ${N_{\text{R}}}$ receiving antennas $\{ {{\mathbf{y}}_m^{\text{d}}} \}_{m = 1}^{{N_{\text{R}}}}$ are concatenated as a vector ${\mathbf{y}}$, such that ${\mathbf{y}} = {[ {{{( {{\mathbf{y}}_1^{\text{d}}} )}^T},\ldots,{{( {{\mathbf{y}}_{{N_{\text{R}}}}^{\text{d}}} )}^T}} ]^T} \in {\mathbb{C}^{{N_{\text{R}}}( {K - P} ) \times 1}}$. Therefore, the relationship between received signals and transmitted symbols is represented as
\begin{equation}
  {\mathbf{y}} = {\mathbf{Xh}} + {\mathbf{w}},
  \label{equ_y_mimo_ofdm}
\end{equation}
where the transmitted block ${\mathbf{X}} = {{\mathbf{I}}_{{N_{\text{R}}}}} \otimes [ {( {{{\mathbf{1}}_{1 \times {N_{\text{T}}}}} \otimes {{\mathbf{I}}_{( {K - P} )}}} ){\text{diag}}( {\mathbf{x}} )} ] \in {\mathbb{C}^{{N_{\text{R}}}( {K - P} ) \times {N_{\text{R}}}{N_{\text{T}}}( {K - P} )}}$ is derived by the concatenation of ${N_{\text{T}}}$ transmitted blocks ${\mathbf{x}} = [ {{{( {{\mathbf{x}}_1^{\text{d}}} )}^T}, \ldots ,{{( {{\mathbf{x}}_{{N_{\text{T}}}}^{\text{d}}} )}^T}} ]^T \in {\mathbb{C}^{{N_{\text{T}}}( {K - P} ) \times 1}}$. Similarly, the corresponding channel vector ${\mathbf{h}}$ and noise vector ${\mathbf{w}}$ can be represented as ${\mathbf{h}} = [ {{{( {{\mathbf{h}}_{11}^{\text{d}}} )}^T}, \ldots ,{{( {{\mathbf{h}}_{1{N_{\text{T}}}}^{\text{d}}} )}^T}, \ldots ,{{( {{\mathbf{h}}_{{N_{\text{R}}}{N_{\text{T}}}}^{\text{d}}} )}^T}} ]^T \in {\mathbb{C}^{{N_{\text{R}}}{N_{\text{T}}}( {K - P} ) \times 1}}$ and ${\mathbf{w}} = [ {{{( {{\mathbf{w}}_1^{\text{d}}} )}^T}, \ldots ,{{( {{\mathbf{w}}_{{N_{\text{R}}}}^{\text{d}}} )}^T}} ]^T \in {\mathbb{C}^{{N_{\text{R}}}( {K - P} ) \times 1}}$, respectively.
\par
Based on the rearrangement of the MIMO-OFDM system, the LMMSE principle can be utilized. Specifically, the following Wiener-Hopf equation is employed \cite{daMJCDKay}

\begin{equation}
    {\hat{\mathbf{h}}_{{\text{LMMSE}}}}  = {{\mathbf{C}}_{{\mathbf{yh}}}^{\mathrm{H}}} {{\mathbf{C}}_{{\mathbf{yy}}}^{-1}} {\mathbf{y}},
  \label{equ_wiener_hopf}
\end{equation}
where ${{\mathbf{C}}_{{\mathbf{yh}}}}$ and ${{\mathbf{C}}_{{\mathbf{yy}}}}$ are defined as 
\begin{equation}
  {{\mathbf{C}}_{{\mathbf{yh}}}} = \mathbb{E}\left\{ {{\mathbf{y}}{{\mathbf{h}}^{\mathrm{H}}}} \right\},
  ~~
  {{\mathbf{C}}_{{\mathbf{yy}}}} = \mathbb{E}\left\{ {{\mathbf{y}}{{\mathbf{y}}^{\mathrm{H}}}} \right\}.
  \label{equ_def_Cyh&Cyy}
\end{equation} 
The derivation involves three types of random variables: the channel coefficient $h$, the transmitted symbol $x$, and the noise variable $w$ according to (\ref{equ_y_mimo_ofdm}). Therefore, we state the following assumptions before computing the second-order moments in (\ref{equ_def_Cyh&Cyy}):
\begin{enumerate}[1)]
  \item The concerned random variables $h$, $x$ and $w$ are mutually independent.
  \item The correlation of channel coefficients, i.e., $\mathbb{E}\{ {{h_i}h_j^ * } \}$, is derived from the second-order statistics of MIMO-OFDM channels.
  \item Different transmitted symbols are mutually independent, i.e., $\mathbb{E}\{ {{x_i}x_j^ * } \} = \mathbb{E}\{ {{x_i}} \}\mathbb{E}\{ {x_j^ * } \}$ for any $i \ne j$.
  \item The signal detection error is defined as $\Delta e = x - \hat x$, where the estimated symbol is obtained by the expectation of symbol, i.e., $\mathbb{E}\{ x \} = \hat x$. That is to say, the estimation of symbols is assumed to be unbiased.
\end{enumerate}
\par
Consequently, the following statistical properties are deduced:
\begin{enumerate}[(a)]
  \item Statistical property of noise variable: $\mathbb{E}\{ {{w_i}} \} = 0$, $\mathbb{E}\{ {{w_i}w_i^ * } \} = \sigma _w^2$, and $\mathbb{E}\{ {{w_i}w_j^ * } \} = 0$ for any $i \ne j$.
  \item Statistical property of detection error: $\mathbb{E}\{ {\Delta {e_i}} \} = 0$, $\mathbb{E}\{ {\Delta {e_i}\Delta e_i^ * } \} = {v_i}$, and $\mathbb{E}\{ {\Delta {e_i}\Delta e_j^ * } \} = 0$ for any $i \ne j$, where ${v_i}$ is the given information in ${\mathbf{V}}_{\text{p}}^{( T )}$ computed by (\ref{ep_Vp_xp}).
  \item Statistical property of transmitted symbol: $\mathbb{E}\{ {{x_i}} \} = {\hat x_i}$, $\mathbb{E}\{ {{x_i}x_i^ * } \} = {{\hat x}_i}\hat x_i^ *  + {v_i}$, and $\mathbb{E}\{ {{x_i}x_j^ * } \} = {{{\hat x}_i}\hat x_j^ * }$ for any $i \ne j$.
\end{enumerate}
According to these properties, the detailed derivation is presented in Appendix A. In general, the deduced form of LMMSE estimation in channel coefficients corresponding to data subcarriers, using estimated symbols derived in the previous JCD layer, is given by \eqref{equ_wiener_hopf}. The formulation is as follows
\vspace{-1em}
\begin{subequations}
\begin{align}
  \label{equ_mimo_ofdm_jcd_lmmse}
   {{\bf{C}}_{{\bf{yh}}}} &= {\hat{\bf{X}}}{{\bf{C}}_{{\bf{hh}}}},  \\
   {{\bf{C}}_{{\bf{yy}}}} &= {\hat{\bf{X}}}{{\bf{C}}_{{\bf{hh}}}}{{\hat{\bf{X}}}^{\mathrm{H}}} + {\bf{AB}}{{\bf{A}}^{\mathrm{H}}} + \sigma _w^2{{\bf{I}}_{{N_{\text{R}}}\left( {K - P} \right)}}
\end{align}
\end{subequations}
with ${\mathbf{A}} = {{\mathbf{I}}_{{N_{\text{R}}}}} \otimes {{\mathbf{1}}_{1 \times {N_{\text{T}}}}} \otimes {{\mathbf{I}}_{K - P}}$ and ${\mathbf{B}} = {{\mathbf{C}}_{{\mathbf{hh}}}} \odot ( {{{\mathbf{1}}_{{N_{\text{R}}} \times {N_{\text{R}}}}} \otimes {\mathbf{V}}} )$. Here, ${{\mathbf{C}}_{{\mathbf{hh}}}}$ refers to the second-order statistics of MIMO-OFDM channels, which can be acquired through ${{\mathbf{C}}_{{\mathbf{hh}}}} = \mathbb{E}\{ {{\mathbf{h}}{{\mathbf{h}}^{\mathrm{H}}}} \} \in {\mathbb{C}^{{N_{\text{R}}}{N_{\text{T}}}( {K - P} ) \times {N_{\text{R}}}{N_{\text{T}}}( {K - P} )}}$. ${\mathbf{V}} = {\text{diag}} ( {{v_1},\ldots,{v_{{N_{\text{T}}}( {K - P} )}}}  )$ refers to the autocorrelation of signal detection error acquired in the EP detector in the previous JCD layer. Specifically, ${v_i} = \mathbb{E}\{ {\Delta {e_i}\Delta e_i^ * } \}$ corresponds to the $i$-th diagonal elements in ${\mathbf{V}}_{\text{p}}^{( T )}$, which is computed according to (\ref{ep_Vp_xp}).

\vspace{-1em}
\subsection{Complexity Analysis of MJCD-LMMSE}
\par
Before conducting experimental realization, the number of layers in the proposed JCD structure, i.e., $I$, is worth discussing, to provide the most effective performance promotion with the least rounds of JCD iteration. It is empirically observed that $I=2$ is preferred, reaching the most efficient improvement compared to the original design, i.e., traditional LMMSE and EP. That is to say, after acquiring the original channel estimates and performing EP detection accordingly, only one extra JCD iteration is needed. The reason for the setting of $I=2$ is discussed in Section \ref{sec_simulation}.
\par
Even though the least number of JCD layers has been adopted to reduce complexity as much as possible, the proposed \emph{MJCD-LMMSE} still involves substantial computational cost. Specifically, the complexity of this method is $\mathcal{O}( {N_{\text{R}}^3N_{\text{T}}^2{{( {K - P} )}^3}} )$. For common MIMO-OFDM configuration, for example, $8 \times 8$ MIMO, $K=256$ and $P=16$, the computational cost can be prohibitive.
\par
The high complexity of the proposed \emph{MJCD-LMMSE} can be intractable, where the cost is dominated by operations of matrix multiplication rather than matrix inversion. According to (\ref{equ_mimo_ofdm_jcd_lmmse}), the computation of ${{\mathbf{C}}_{{\mathbf{yh}}}}$ and ${{\mathbf{C}}_{{\mathbf{yy}}}}$ involve the multiplications of two matrices with dimensions of ${{N_{\text{R}}}( {K - P} ) \times {N_{\text{R}}}{N_{\text{T}}}( {K - P} )}$ and ${{N_{\text{R}}}{N_{\text{T}}}( {K - P} ) \times {N_{\text{R}}}{N_{\text{T}}}( {K - P} )}$, respectively. Therefore, common algorithms that substitute iterative approximation for direct matrix inversion, for example, Gauss-Seidel method \cite{complexGSDai}, are not suitable for the target of reducing complexity under such circumstances.
\par
On the other hand, there is an opportunity for further exploration into methods for reducing complexity in the derivation process of the proposed approach. Given the notable computational overhead associated with matrix multiplications, careful attention to dimensionality reduction is warranted. Fundamentally, the utilization of the LMMSE principle within OFDM subsystems emerges as a potential solution. This approach eliminates the requirement for large-dimension equivalence and serves to mitigate concerns surrounding computational costs in matrix multiplications. A detailed discussion of this concept is provided in the next section.

\section{Low-Complexity Equivalent of Proposed Data-aided Method}
\label{sec_OFDM_CE}
\par
In this section, another equivalent representation of LMMSE-based data-aided method is proposed, namely, \emph{OJCD-LMMSE}, aiming at reducing computational complexity. The proposed equivalent algorithm is deduced in separate OFDM subsystems, which mitigates the high-dimensional computations involved in the \emph{MJCD-LMMSE} algorithm.
\vspace{-1em}
\subsection{OJCD-LMMSE}

\par
The orthogonality principle for LMMSE-based channel estimation in OFDM systems
has been applied in the traditional estimator in JCD-1 as shown in (\ref{equ_h_mn_LS})-(\ref{equ_h_mn_LMMSE}). Similarly, as for application in the second JCD layer, preliminary LS estimation at data subcarriers is
\begin{equation}
  {\mathbf{h}}_{m,n}^{{\text{LS,new}}} = {\left( {{\mathbf{X}}_n^{\text{d}}} \right)^{ - 1}}{\mathbf{y}}_{m,n}^{\text{d}}.
  \label{equ_h_mn_LSnew1}
\end{equation}
However, the computation in (\ref{equ_h_mn_LSnew1}) analogous to (\ref{equ_h_mn_LS}) is intractable, not only because the transmitted data symbols are unknown for detection, but also because the output at the $m$-th receiving antenna corresponding to the $n$-th transmitted data block ${\mathbf{y}}_{m,n}^{\text{d}}$ is not available. Instead, the $m$-th received component corresponding to the $n$-th transmitted data block can be estimated \cite{daCMIMOOFDMYli} through
\begin{equation}
  {\mathbf{\hat y}}_{m,n}^{\text{d}} = {\mathbf{y}}_m^{\text{d}} - \sum\limits_{\substack{n' = 1\\n' \ne n}}^{{N_{\text{T}}}} {{\hat{\mathbf{X}}}_{n'}^{\text{d}}\hat{\mathbf{h}}_{m,n'}^{\text{d}}},
  \label{equ_yhat_mn_d}
\end{equation}
under which circumstance the data estimates error should be considered in the derivation of LMMSE channel estimates. A principle is presented in \cite{daCOMPKas} for measuring the influence of symbol error through a weighting matrix. However, the decision of the matrix elements is made through a fixed threshold and is exclusively deduced for BEM-based methods under the Gaussian assumption of expansion coefficients. Therefore, this scheme does not apply to the general form of LMMSE estimates, let alone the potential inaccuracy induced by the nonuse of true data error information.
\par
In our proposed method, the estimated received component is equivalently represented as
\begin{equation}
  {\mathbf{\hat y}}_{m,n}^{\text{d}} = {\hat{\mathbf{X}}}_n^{\text{d}}{\mathbf{h}}_{m,n}^{\text{d}} + {{\mathbf{Z}}_{m,n}},
\end{equation}
where ${{\mathbf{Z}}_{m,n}}$ includes all interference and is treated as the equivalent noise term
\begin{equation}
  {{\mathbf{Z}}_{m,n}} = \sum\limits_{\substack{n' = 1\\n' \ne n}}^{{N_{\text{T}}}} {{\hat{\mathbf{X}}}_{n'}^{\text{d}}{\mathbf{\Delta h}}_{m,n'}^{\text{d}}}  + \sum\limits_{n' = 1}^{{N_{\text{T}}}} {{\mathbf{\Delta }}{{\mathbf{E}}_{n'}}{\mathbf{h}}_{m,n'}^{\text{d}}}  + {\mathbf{w}}_m^{\text{d}},
  \label{equ_Z_mn}
\end{equation}
among which ${{\mathbf{\Delta h}}_{m,n'}^{\text{d}}}$ and ${{\mathbf{\Delta }}{{\mathbf{E}}_{n'}}}$ denote the channel estimation error and signal detection error, respectively
\begin{subequations}
  \begin{align}
    {{\mathbf{\Delta h}}_{m,n'}^{\text{d}}} &= {\mathbf{h}}_{m,n'}^{\text{d}} - \hat{\mathbf{h}}_{m,n'}^{\text{d}}, \\
    {\mathbf{\Delta }}{{\mathbf{E}}_{n'}} &= {\mathbf{X}}_{n'}^{\text{d}} - {\hat{\mathbf{X}}}_{n'}^{\text{d}}.
  \end{align}
  \label{equ_def_error_ofdm}
\end{subequations}
Consequently, the LS computation in (\ref{equ_h_mn_LSnew1}) is now replaced by
\begin{equation}
    {\mathbf{h}}_{m,n}^{{\text{LS,new}}} = {\left( {{\hat{\mathbf{X}}}_n^{\text{d}}} \right)^{ - 1}}{\mathbf{\hat y}}_{m,n}^{\text{d}}
    = {\mathbf{h}}_{m,n}^{\text{d}} + {\left( {{\hat{\mathbf{X}}}_n^{\text{d}}} \right)^{ - 1}}{{\mathbf{Z}}_{m,n}},
  \label{equ_h_mn_LSnew2}
\end{equation}
and the LMMSE estimation is denoted by
\begin{equation}
  {\mathbf{h}}_{m,n}^{{\text{LMMSE,new}}} = {\mathbf{W}}_{{\text{LMMSE}}}^{{\text{new}}}{\mathbf{h}}_{m,n}^{{\text{LS,new}}},
  \label{equ_h_mn_LMMSEnew}
\end{equation}
which requires the derivation of weight ${\mathbf{W}}_{{\text{LMMSE}}}^{{\text{new}}}$ defined as follows
\begin{subequations}
  \begin{align}
    {\mathbf{W}}_{{\text{LMMSE}}}^{{\text{new}}} &= {{\mathbf{R}}_{{{\mathbf{h}}^{{\text{d}}}}{{\mathbf{h}}^{{\text{LS,new}}}}}}{\mathbf{R}}_{{{\mathbf{h}}^{{\text{LS,new}}}}{{\mathbf{h}}^{{\text{LS,new}}}}}^{ - 1}, \\
    {{\mathbf{R}}_{{{\mathbf{h}}^{{\text{d}}}}{{\mathbf{h}}^{{\text{LS,new}}}}}} &= \mathbb{E}\left\{ {{\mathbf{h}}_{m,n}^{\text{d}}{{\left( {{\mathbf{h}}_{m,n}^{{\text{LS,new}}}} \right)}^{\mathrm{H}}}} \right\}, \\
    {{\mathbf{R}}_{{{\mathbf{h}}^{{\text{LS,new}}}}{{\mathbf{h}}^{{\text{LS,new}}}}}} &= \mathbb{E}\left\{ {\left( {{\mathbf{h}}_{m,n}^{{\text{LS,new}}}} \right){{\left( {{\mathbf{h}}_{m,n}^{{\text{LS,new}}}} \right)}^{\mathrm{H}}}} \right\}.
  \end{align}
  \label{equ_W_LMMSEnew}
\end{subequations}
The proposed algorithm is summarized in Algorithm \ref{algo_ofdm_jcd_lmmse}, and the detailed derivation process is written in Appendix B. Note that ${{\mathbf{V}}_n} = {\text{diag}}( {{v_1},\ldots,{v_{K - P}}} )$ refers to the autocorrelation of signal detection error at the $n$-th transmitted data block acquired in EP detector in the first layer. Similarly, ${v_i} = \mathbb{E}\{ {\Delta {e_i}\Delta e_i^ * } \}$ corresponds to the $i$-th diagonal elements in ${\mathbf{V}}_{\text{p}}^{( T )}$, which is acquired through (\ref{ep_Vp_xp}).

\begin{algorithm}[t]
  \SetAlgoLined
  \caption{Low-Complexity \emph{OJCD-LMMSE} Estimator for JCD-2}
  \label{algo_ofdm_jcd_lmmse}
  \KwIn{${{\mathbf{C}}_{{\mathbf{hh}}}}$, $\{ {{\mathbf{y}}_m^{\text{d}}} \}_{m = 1}^{{N_{\text{R}}}}$, $\{ {\hat{\mathbf{h}}_{m,n}^{\text{d}}} \}_{m,n = 1,1}^{{N_{\text{R}}},{N_{\text{T}}}}$, $\{ {{\hat{\mathbf{X}}}_n^{\text{d}}} \}_{n = 1}^{{N_{\text{T}}}}$, $\{ {{\mathbf{X}}_n^{\text{p}}} \}_{n = 1}^{{N_{\text{T}}}}$, $\{ {{{\mathbf{V}}_n}} \}_{n = 1}^{{N_{\text{T}}}}$, ${{\mathbf{W}}_{{\text{LMMSE}}}}$ in JCD-1, $\sigma_w^2$}  \vspace{0.15cm}
  \KwOut{$\{ {{\mathbf{h}}_{m,n}^{{\text{LMMSE,new}}}} \}_{m,n = 1,1}^{{N_{\text{R}}},{N_{\text{T}}}}$} \vspace{0.15cm}
  \SetKw{KwIni}{Initialize:}
  \KwIni Compute ${{\mathbf{R}}_{{{\mathbf{h}}^{\text{d}}}{{\mathbf{h}}^{\text{d}}}}}\left( {{n_1},{n_2}} \right)$, ${{\mathbf{R}}_{{{\mathbf{h}}^{\text{d}}}{{\mathbf{h}}^{\text{p}}}}}\left( {{n_1},{n_2}} \right)$ and ${{\mathbf{R}}_{{{\mathbf{h}}^{\text{p}}}{{\mathbf{h}}^{\text{p}}}}}\left( {{n_1},{n_2}} \right)$ from ${{\mathbf{C}}_{{\mathbf{hh}}}}$ according to (\ref{deriv_def_Rs}), where ${n_1},{n_2} \in \left\{ {1, \ldots ,{N_{\text{T}}}} \right\}$.
  \BlankLine
  Compute ${{\mathbf{W}}_1} = {\mathbf{W}}_{{\text{LMMSE}}}^{\text{d}}$; \\
  Compute ${{\mathbf{V}}_{\text{D}}} = {{\mathbf{R}}_{{{\mathbf{h}}^{\text{d}}}{{\mathbf{h}}^{\text{d}}}}} \odot \sum\limits_{n = 1}^{{N_{\text{T}}}} {{{\mathbf{V}}_n}}$; \\
  Compute ${{\mathbf{V}}^{\text{x}}}\left( {{n_1},{n_2}} \right) = {\hat{\mathbf{x}}}_{{n_1}}^{\text{d}}{\left( {{\hat{\mathbf{x}}}_{{n_2}}^{\text{d}}} \right)^{\mathrm{H}}}$; \\
	\For{$n = 1, \ldots ,{N_{\text{T}}}$}{
 		Compute ${{\mathbf{B}}_n}$ according to (\ref{deriv_Rhhls_step3}); \\
    Compute ${\bm{\Sigma} _n}$ according to (\ref{deriv_Rhlshls_step3}); \\
    Compute ${{\mathbf{R}}_{{{\mathbf{h}}^{{\text{d}}}}{{\mathbf{h}}^{{\text{LS,new}}}}} \left( n \right)}$ according to (\ref{deriv_Rhhls_step1}); \\
    Compute ${{\mathbf{R}}_{{{\mathbf{h}}^{{\text{LS,new}}}}{{\mathbf{h}}^{{\text{LS,new}}}}} \left( n \right)}$ according to (\ref{deriv_Rhlshls_step1}); \\
    Calculate the weight matrix: ${\mathbf{W}}_{{\text{LMMSE}}}^{{\text{new}}} \left( n \right) = {{\mathbf{R}}_{{{\mathbf{h}}^{{\text{d}}}}{{\mathbf{h}}^{{\text{LS,new}}}}} \left( n \right)}{{\mathbf{R}}_{{{\mathbf{h}}^{{\text{LS,new}}}}{{\mathbf{h}}^{{\text{LS,new}}}}}^{ - 1} \left( n \right)}$; \\
    \For{$m = 1, \ldots ,{N_{\text{R}}}$}{
      Compute ${\mathbf{\hat y}}_{m,n}^{\text{d}}$ according to (\ref{equ_yhat_mn_d}); \\
      Perform LS estimation ${\mathbf{h}}_{m,n}^{{\text{LS,new}}}$ according to (\ref{equ_h_mn_LSnew2}); \\
      Perform LMMSE estimation: ${\mathbf{h}}_{m,n}^{{\text{LMMSE,new}}} = {{\mathbf{W}}_{{\text{LMMSE}}}^{{\text{new}}}\left( n \right)} {\mathbf{h}}_{m,n}^{{\text{LS,new}}}$.
    }
	}
\end{algorithm}

\subsection{Complexity Comparison}

\par
With the acquisition of two proposed equivalent representations of the LMMSE-based data-aided method, namely \emph{MJCD-LMMSE} and \emph{OJCD-LMMSE}, it is essential to conduct a comparative analysis of the computational complexity. The comparison using $\mathcal{O} ( \cdot )$ notation is considered. Furthermore,  for a more intuitive comparison, the floating point operations (FLOPs) \footnote{We adopt the widely used definition of FLOPs as the number of multiply-add operations.} involved in the realizations of both algorithms are considered. According to Table \ref{tab_complexity1}, it is obvious that the realization of \emph{MJCD-LMMSE} suffers from high computational cost at the $4 \times 4$, $K=128$, and $P=8$ MIMO-OFDM configuration. In contrast, substituting the \emph{MJCD-LMMSE} estimator with the \emph{OJCD-LMMSE} estimator at JCD-2 effectively reduces the FLOPs by three orders of magnitude.
\begin{table}[t] 
  \caption{Complexity Analysis of proposed algorithms}
  \label{tab_complexity1}
  \centering
  \begin{tabular}{ccc}
    \toprule
    Algorithm & FLOPs & Complexity\\
    \midrule
    \makecell[c]{\emph{MJCD-LMMSE}} & $2.66 \times 10^9$      & $\mathcal{O}( {N_{\text{R}}^3N_{\text{T}}^2{{( {K - P} )}^3}} )$ \\
    \makecell[c]{\emph{OJCD-LMMSE}} & $4.84\times 10^6$  & $\mathcal{O}( {P{N_{\text{T}}}{{( {{N_{\text{T}}} - 1} )}^2}{{( {K - P} )}^2}} )$ \\
    \bottomrule
  \end{tabular}
  \vspace{-1em}
\end{table}

\section{Simulation Results}
\label{sec_simulation}
\par
In this section, numerical results are presented. The detailed parameter settings in the MIMO-OFDM system are first shown. Subsequently, discussions on convergence and equivalence of the proposed JCD structure are presented. Moreover, the impact of pilot lengths is investigated. Finally, the bit error rate (BER) performance under various scenarios is evaluated.

\vspace{-1em}
\subsection{Parameter Settings}
\par
\begin{figure}[!t]
  \centering
  \subfloat[{Block-fading.}]{
    \includegraphics[scale=0.49]{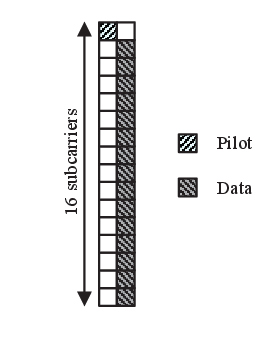}}
  \subfloat[{Time-varying, $N_\text{p}$ OFDM symbols are selected \\ for pilot transmission.}]{
    \includegraphics[scale=0.49]{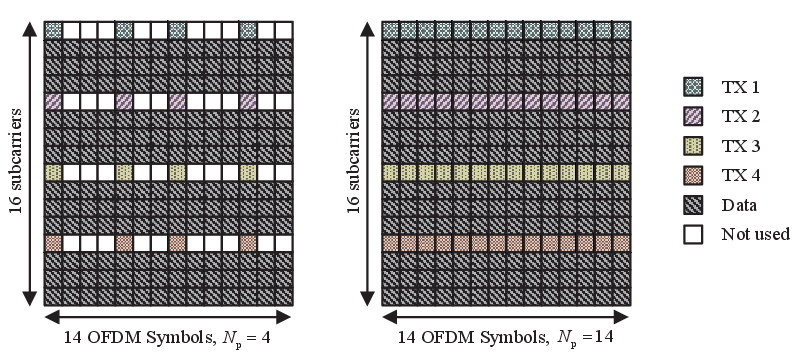}}
  \caption{~Frame structures under different scenarios.}
  \label{fig_pilot_pattern}
\end{figure}

A MIMO-OFDM system configured with $4 \times 4$ or $8 \times 8$ antennas and $K=128$ or $256$ subcarriers is considered. The modulation type for transmitted data streams is set as quadrature phase-shift keying (QPSK) or 16-QAM. Both block-fading and time-varying scenarios are implemented, utilizing different frame structures as illustrated in Fig. \ref{fig_pilot_pattern}. For evaluations of performance improvements based on the proposed algorithms, the frame design depicted in Fig. \ref{fig_pilot_pattern}(a), consistent with the assumption in Section \ref{sec_system_model}, is utilized in Sections \ref{sec_simulation}-B through \ref{sec_simulation}-D. Conversely, for the performance assessment of the proposed receiver under various time-selective fading environments, the pilot pattern shown in Fig. \ref{fig_pilot_pattern}(b) is employed. Note that the figure takes the $4 \times 4$ with $K=128$, $P=8$ case as an example, and uses the corresponding pilot interval to represent the resource block. The number of JCD layers, as mentioned above, is $I=2$. Besides, in the detection module, the number of EP iterations is fixed at $T=5$, and the damping factor value is empirically set as $\beta  = 0.2$.

\par
The tapped delay line (TDL) channel model is adopted during the generation of MIMO-OFDM channel datasets, where TDL-C, the non-line-of-sight (NLOS) case profile with 24 taps, is utilized. The correlation between antennas is also taken into consideration. Specifically, the Kronecker model is adopted for the representation of MIMO characteristic, i.e., ${{\mathbf{H}}_{{\text{corr}}}} = {\mathbf{R}}_{\text{r}}^{1/2}{{\mathbf{H}}_{{\text{iid}}}}{\mathbf{R}}_{\text{t}}^{1/2}$, where ${\mathbf{R}}_{\text{t}}$ and ${\mathbf{R}}_{\text{r}}$ denotes the spatial correlation of transmitter and receiver, which is illustrated by a correlation coefficient $\rho $ using exponential correlation model such that the element of matrices ${r_{ij}}$ fulfills
\begin{equation}
  {r_{ij}} = \left\{ {\begin{array}{*{20}{c}}
    {\rho ^{j - i}}, &i \leqslant j, \\
    r_{ji}^ *, &i > j.
  \end{array}} \right.
\end{equation}
The corresponding spatial matrices are utilized for assignment in nrTDLChannel \footnote{The MATLAB 5G Toolbox function.}. The configurations concerned for MIMO-OFDM system and channel generation are summarized in Table \ref{tab_mimo_ofdm}. Note that $\rho=0$ is used for channel generation unless noted otherwise.
\begin{table}[t] 
  \caption{Parameters for MIMO-OFDM Simulation}
  \label{tab_mimo_ofdm}
    \centering
    \begin{tabular}{ll}
     \toprule
     Parameter & Value \\
     \midrule
     Antennas & $4 \times 4,\ 8 \times 8$ \\
     Subcarriers & $128,\ 256$ \\
     Pilots & $8,\ 16,\ 32$ \\
     Modulation & QPSK, \ 16-QAM \\
     JCD Layers & $2$ \\
     EP Iterations & $5$ \\
     Damping Factor & $0.2$ \\
     Delay Profile & TDL-C (NLOS) \\
     Carrier Frequency & $3.5$ GHz \\
     Delay Spread & $200$ ns \\
     Subcarrier Spacing & $15$ kHz \\
     MIMO Correlation & $\rho  = 0, \ 0.5$ \\
  \bottomrule
  \end{tabular}
\end{table}

\vspace{-1em}
\subsection{Layers in MJCD-LMMSE Based JCD Structure}
\par
The high complexity of the proposed \emph{MJCD-LMMSE} makes it necessary to avoid extra JCD iterations, such that the computational cost is reduced as much as possible. For discussions on the convergence of the \emph{MJCD-LMMSE}-based JCD design, $4 \times 4$ MIMO with $K=128$ and $P=16$ is configured. Both estimation accuracy of ${\hat{\mathbf{h}}_{{\text{LMMSE}}}}$ and detection error of ${\hat{\mathbf{x}}}$ are evaluated under the JCD setting at $I=2$ and $I=5$, respectively. The case of the traditional method, i.e., $I=1$, is utilized as the benchmark under different signal-to-noise ratios (SNRs) so that the extent of performance promotion after different JCD iterations is observed intuitively. Notably, the channel estimation accuracy is measured by the mean square error (MSE)
\begin{equation} 
  {\text{MSE}} = \frac{1}{{2{N_{{\text{test}}}}{N_{\text{R}}}{N_{\text{T}}}\left( {K - P} \right)}}\sum\limits_{i = 1}^{{N_{{\text{test}}}}} {{{\left\| {\hat{\mathbf{h}}_i - {\mathbf{h}}_i} \right\|}^2}},
\end{equation}
where ${\mathbf{h}}_i$ denotes the MIMO-OFDM channels corresponding to data transmission during the $i$-th test, and ${{N_{{\text{test}}}}}$ is the number of testing realizations. The signal detection error is evaluated by BER.
\par
\begin{figure}[t]
  \begin{minipage}{3.5in}
    \centerline{\includegraphics[width=3.5in]{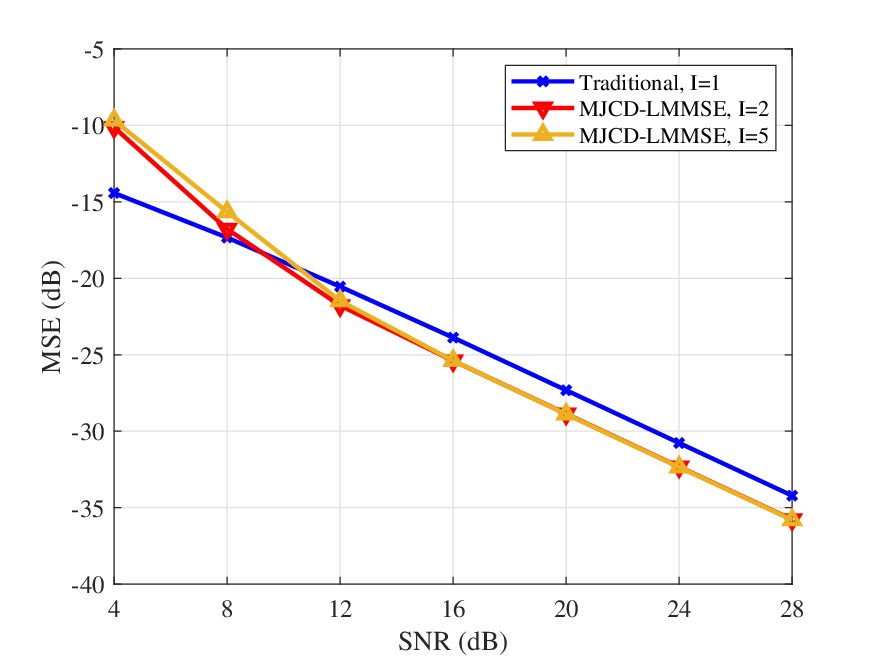}}
    \centerline{ (a) MSE of channel estimation}
  \end{minipage}
  \hfill
  \begin{minipage}{3.5in}
    \centerline{\includegraphics[width=3.5in]{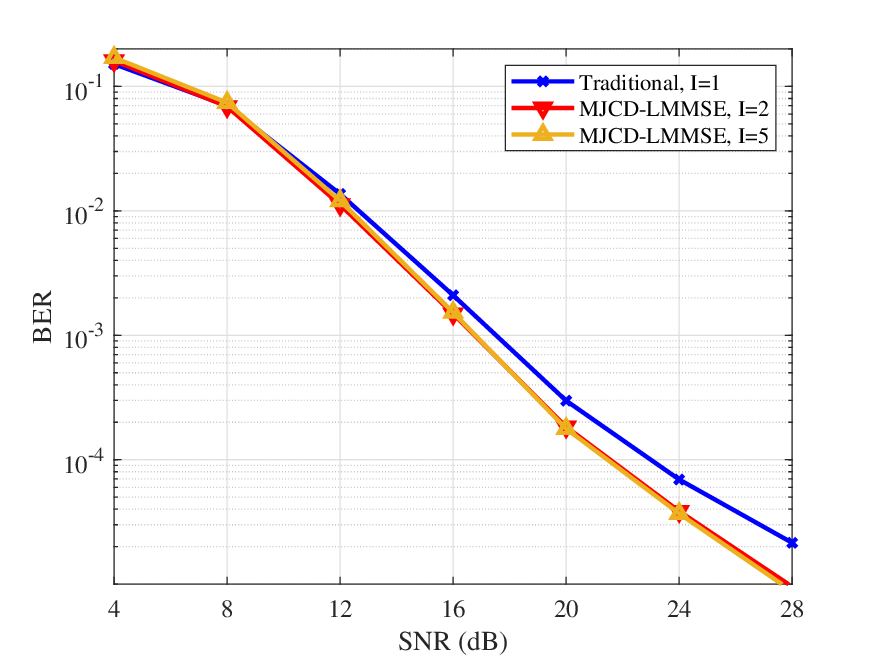}}
    \centerline{ (b) BER of signal detection}
  \end{minipage}
  \caption{~Accuracy of channel estimation and signal detection at the output of JCD structure under different settings of $I$.}
  \label{fig_converg_qpsk}
  \vspace{-1em}
\end{figure}

Fig. \ref{fig_converg_qpsk} shows the performance comparison under QPSK modulation. From the figure, it is obvious that the gains in performance occur in JCD-2, while from JCD-3 to JCD-5, the performance tends to saturate. Moreover, according to Fig. \ref{fig_converg_qpsk}(a), under $\text{SNR} \in [ {4,\, 8} ] \ \text{dB}$, the use of JCD design may lead to negative effects on estimation accuracy. This is reasonable, for the detection accuracy is poor in low SNRs according to Fig. \ref{fig_converg_qpsk}(b), thus the estimated symbol ${\hat{\mathbf{X}}}$ used in \emph{MJCD-LMMSE} is inaccurate, which may deviate from the assumption of unbiased estimation according to Section \ref{sec_MIMO_OFDM_CE}-A, 4). Despite the lower estimation accuracy compared to $I=1$ in low SNRs, the corresponding detection performance does not visibly degrade, as shown in Fig. \ref{fig_converg_qpsk}(b). As for $\text{SNR} \in [ {12,\,28} ] \ \text{dB}$, the improvement of both estimation and detection accuracy can be observed in JCD-2. The comparisons under 16-QAM modulation showcase similar results, which are not exhibited. In general, it is empirically concluded that $I=2$ provides the most efficient improvement, which is implemented in the rest of the simulation.

\vspace{-1em}
\subsection{Performance Comparison Between MJCD-LMMSE and OJCD-LMMSE}
\par
In this subsection, we evaluate the performance comparison between the proposed \emph{OJCD-LMMSE} and \emph{MJCD-LMMSE}. Theoretically, equivalence in performance is expected, as \emph{OJCD-LMMSE} is derived equivalently using the LMMSE principle in OFDM subsystems. To further validate this, the following simulation is organized:
Similar to the previous subsection, a MIMO-OFDM system configured with $4 \times 4$ antennas and $K=128$, $P=16$ is considered, and the detection error at the output of JCD-2 under different SNRs is evaluated, where the proposed algorithms are utilized at JCD-2 for performance comparison. Besides, the case of $I=1$, i.e., traditional LMMSE and EP, is utilized as the baseline.

\begin{figure}[t]
  \begin{minipage}{3.5in}
    \centerline{\includegraphics[width=3.5in]{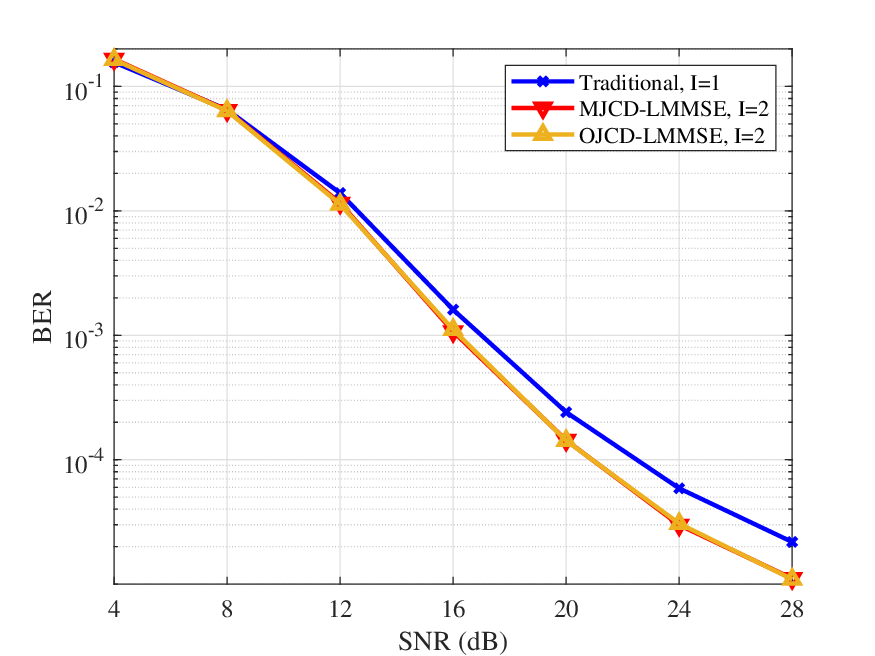}}
  \end{minipage}
  \caption{~Detection performance comparison of JCD design using proposed algorithms.}
  \label{fig_equival}
  \vspace{-1em}
\end{figure}
\par
Fig. \ref{fig_equival} shows the comparison of the proposed methods under QPSK modulation. On the one hand, both of the JCD designs adopting the proposed algorithms show significant improvement compared to the benchmark with traditional algorithms. On the other hand, \emph{OJCD-LMMSE} provides comparable performance to that of \emph{MJCD-LMMSE} based JCD structure. In summary, simulation results in Fig. \ref{fig_equival} further validate the performance equivalence of our proposed designs, yet \emph{OJCD-LMMSE} based JCD structure significantly reduces the computational cost. Therefore, \emph{OJCD-LMMSE}-based JCD structure is adopted in the remainder of the performance comparisons.

\vspace{-1em}
\subsection{Influences of Pilot Information}
\par
Another factor that influences the performance of JCD remains for discussion, which is the length of inserted pilot sequences for preliminary estimation in JCD-1. According to (\ref{equ_h_mn_LS})-(\ref{equ_W_LMMSE}), using more subcarriers for inserting pilots naturally leads to more accurate channel estimation with traditional LMMSE, as more information becomes available. Consequently, it is predictable that if the number of inserted pilots is sufficient, utilizing the JCD structure to facilitate more reliable EP detection becomes unnecessary since traditional LMMSE can already provide accurate estimation results. Conversely, the fewer the pilots inserted, the more significant the performance improvement offered by the proposed JCD structure compared to the baseline scenario where $I=1$. To provide empirical evidence for the aforementioned analysis, the following simulation is conducted.

\subsubsection{Performance Under Different Pilot Sequence Lengths}
Consider both of the MIMO-OFDM configurations, that is, $4 \times 4$ MIMO with $K=128$ and $8 \times 8$ MIMO with $K=256$. Three different types of pilot settings, i.e., $P \in \{ 8,16,32 \}$ are considered. Under each scenario both the detection accuracy of $I=1$ and $I=2$ are presented. Moreover, the performance of EP detector when perfect CSI is available is also shown as a reference point.
\begin{figure}[t]
  \begin{minipage}{3.5in}
    \centerline{\includegraphics[width=3.5in]{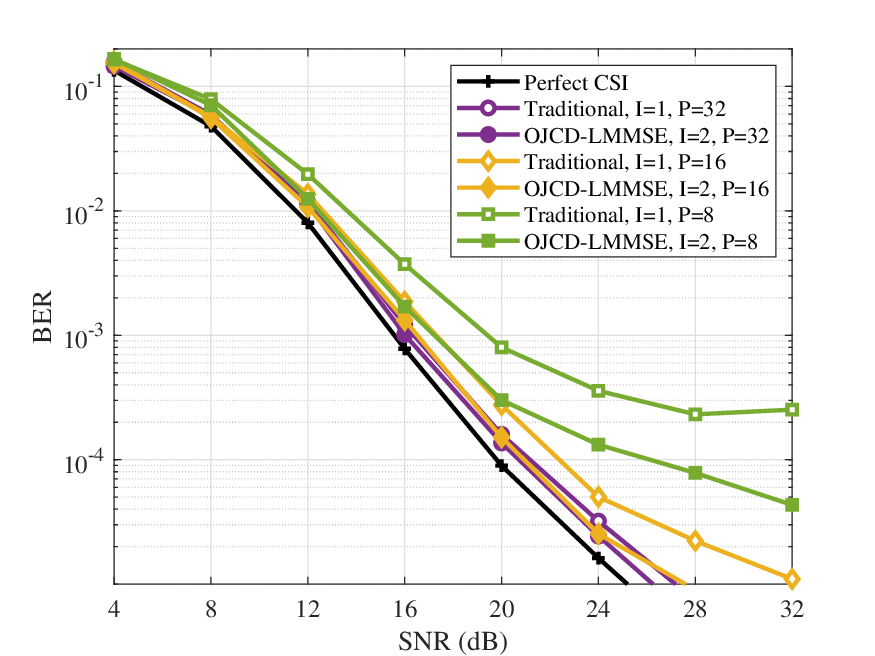}}
    \centerline{(a) BER under QPSK modulation}
  \end{minipage}
  \hfill
  \begin{minipage}{3.5in}
    \centerline{\includegraphics[width=3.5in]{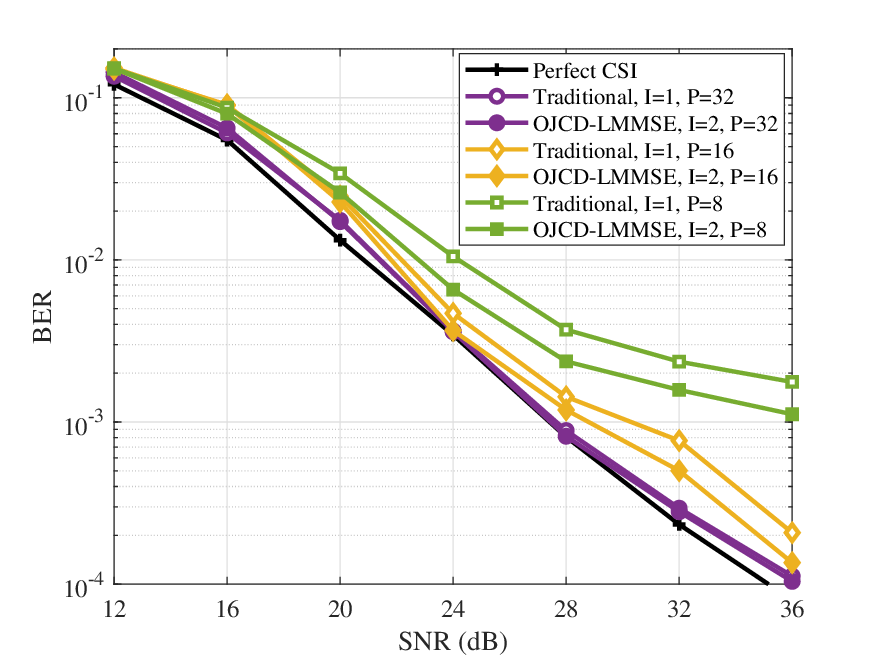}}
    \centerline{(b) BER under 16-QAM modulation}
  \end{minipage}
  \caption{~Detection performance of the proposed \emph{OJCD-LMMSE} with different pilot lengths under $4 \times 4$ MIMO with $K=128$.}
  \vspace{-1em}
  \label{fig_pilot_4}
\end{figure}
\par
Fig. \ref{fig_pilot_4} shows the performance comparison under different SNRs in MIMO-OFDM system with $4 \times 4$ antennas and $128$ subcarriers. According to Fig. \ref{fig_pilot_4}(a), it is observed that our proposed JCD structure provides remarkable gains when $P=8$ and significantly improves the performance saturation caused by inaccurate channel estimation under high SNR regions. As for $P=16$, a performance gain of over $4$ dB is achieved. A similar phenomenon can be observed for 16QAM modulation, as shown in Fig. \ref{fig_pilot_4}(b), where $P=16$ provides a gain of approximately $2$ dB compared to $I=1$. On the contrary, when $P=32$, even though performance gain still exists in Fig. \ref{fig_pilot_4}(a), the gain is less pronounced, mainly because the traditional design provides accurate channel estimation and consequently precise signal detection, leaving little room for improvement, as we refer to the ideal case denoted by Perfect CSI. In Fig. \ref{fig_pilot_4}(b), the accuracy even tends to converge at $I=1$.
\begin{figure}[t]
  \begin{minipage}{3.5in}
    \centerline{\includegraphics[width=3.5in]{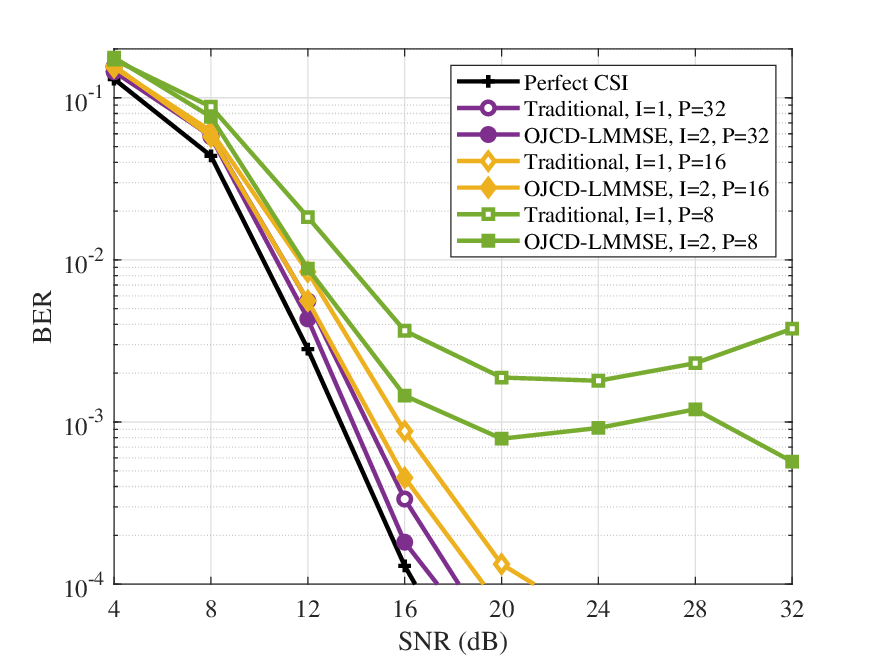}}
    \centerline{(a) BER under QPSK modulation}
  \end{minipage}
  \hfill
  \begin{minipage}{3.5in}
    \centerline{\includegraphics[width=3.5in]{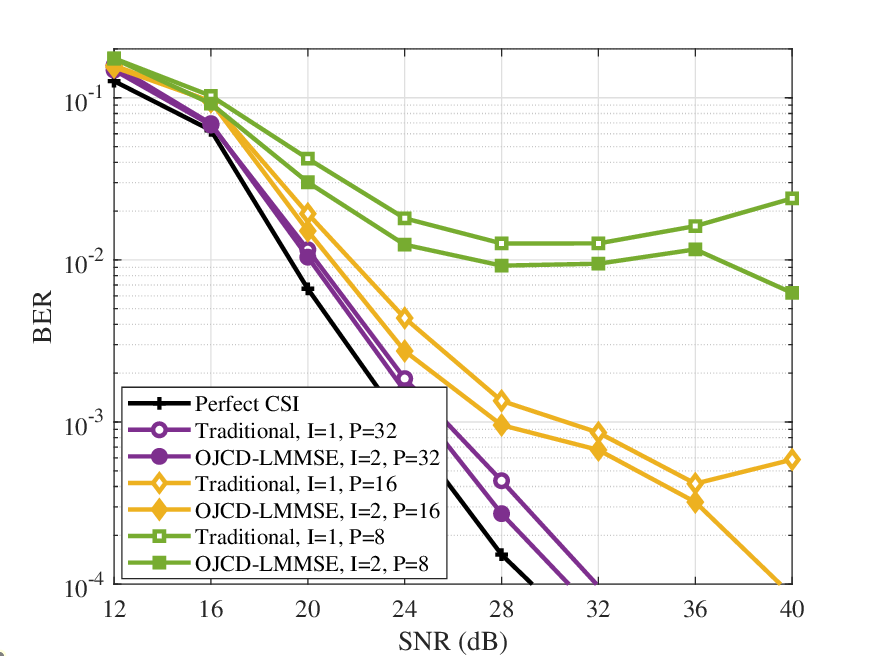}}
    \centerline{(b) BER under 16-QAM modulation}
  \end{minipage}
  \caption{~Detection performance of the proposed \emph{OJCD-LMMSE} with different pilot lengths under $8 \times 8$ MIMO with $K=256$.}
  \vspace{-1em}
  \label{fig_pilot_8}
\end{figure}
\par
The same comparisons are conducted in a MIMO-OFDM system with $8 \times 8$ antennas and $256$ subcarriers, as shown in Fig. \ref{fig_pilot_8}. Under QPSK modulation, according to Fig. \ref{fig_pilot_8}(a), similar findings are observed, that is, the fewer pilots inserted, the more improvement our proposed JCD provides. Notably, for $P=8$, despite the apparent performance promotion compared to $I=1$, the proposed JCD structure exhibits no enhancement of detection accuracy with the increase of SNR when $\text{SNR} \in [ {20,\, 28} ] \text{dB}$. The reason for such a problem, in simple terms, is that the estimated channel coefficients significantly deviate from the actual CSI in the abovementioned SNR region, making it difficult for EP detection to approximate the true posterior distribution in (\ref{equ_p_x_y}) \cite{tradMIMOSDGha}. Similar trends are observed for 16QAM modulation, as shown in Fig. \ref{fig_pilot_8}(b).
\par
In general, the proposed JCD shows substantial enhancement under inadequate pilot information.
The setting of $P=16$ presents a desirable compromise between detection accuracy and the extent of improvement, making it a reasonable choice for the JCD structure. Furthermore, to further explore the potential of our proposed \emph{OJCD-LMMSE} when $P=8$, the decoder is introduced at the output of the JCD structure to compensate for performance saturation, as discussed below.

\subsubsection{Incorporating Channel Decoder}
According to the $8 \times 8$ case in Fig. \ref{fig_pilot_8}, degradation or saturation caused by inadequate pilot information when $P=8$ restricts the performance of our proposed JCD structure. Therefore, for a further evaluation of the performance enhancement, the Viterbi decoder is introduced according to Fig. \ref{fig_jcd_structure}. Specifically, the decoder utilizes the extrinsic information $( {{\mathbf{x}}_{\text{e}}^{( T )},{\mathbf{V}}_{\text{e}}^{( T )}} )$ in (\ref{ep_Ve_xe}) output by EP detector at JCD-$I$, and computes the corresponding pdf ${p_{\text{e}}}( {{\mathbf{x}}|{\mathbf{y}}} ) \sim \mathcal{N}( {{\mathbf{x}}:{\mathbf{x}}_{\text{e}}^{( T )},{\mathbf{V}}_{\text{e}}^{( T )}} )$, which is subsequently demapped as extrinsic LLRs and delivered for channel decoding \cite{decodEPZhang}.
\par
\begin{figure}[t]
  \begin{minipage}{3.5in}
    \centerline{\includegraphics[width=3.5in]{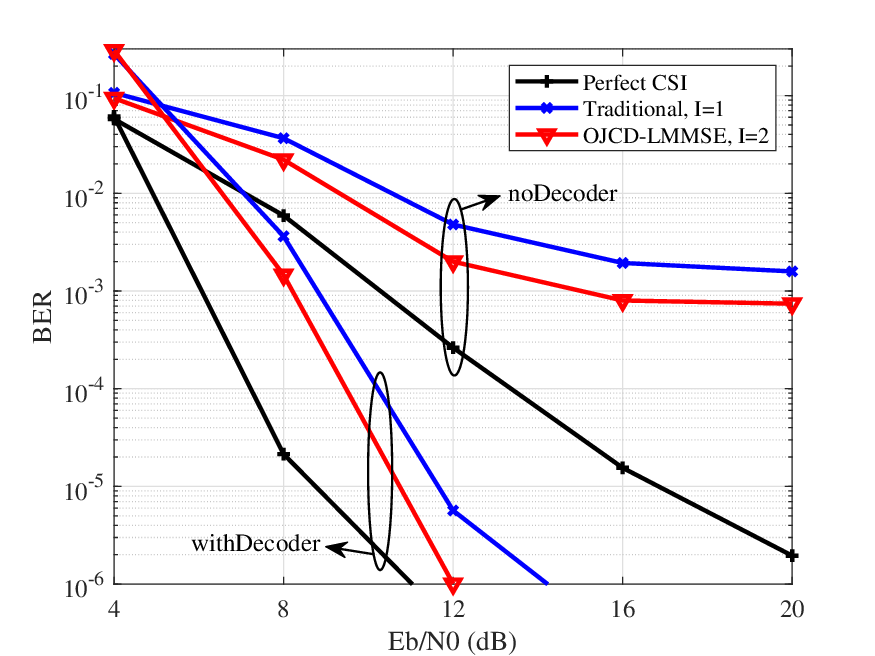}}
  \end{minipage}
  \caption{~BER performance of the proposed \emph{OJCD-LMMSE} under $8 \times 8$ MIMO with $K=256, P=8$.}
  \label{fig_decoding_8}
  \vspace{-1em}
\end{figure}

Fig. \ref{fig_decoding_8} shows the performance with and without the decoding module, where QPSK modulation with $(1984,1/2)$ convolutional code is considered for the coded case. Note that we use $E_b / N_0$ as metric. Moreover, to ensure fairness, the $E_b / N_0$ of the uncoded case is 3 dB more than that of the coded case. As shown in Fig. \ref{fig_decoding_8}, our proposed JCD design exhibits better performance than the traditional design, whether channel decoding is considered or not. Furthermore, the tendency for performance saturation due to inadequate pilot information in the uncoded scenario is effectively mitigated upon incorporating the decoder. Consequently, the performance disparity between the ideal situation and our proposed JCD scheme is reduced. Overall, our proposed JCD receiver delivers satisfactory performance with minimal pilot redundancy.

\vspace{-1em}
\subsection{Performance Comparisons with BEM-based Receiver}
\par
The introduction of basis functions in \cite{daCOMPKas} provides another baseline for conducting LMMSE-based data-aided estimation, using $M_t$ B-Spline functions \cite{daCOMPZak, daCOMPKha} and $M_f$ DPS sequences \cite{daCOMPHam} to represent the time-frequency characteristics of fading channels. A weighting principle (i.e., \emph{WLMMSE}) is designed under BEM to measure the influence of data detection error, which is also deployed in our proposed system. In this section, simulations are conducted under different time-varying scenarios to evaluate whether our proposed JCD receiver reaches comparable accuracy to that of BEM-based receiver, and complexity analysis is investigated.
\par

The $4 \times 4$ MIMO system with $K=128$ subcarriers is configured, and the spatial correlation coefficient is set to $\rho=0.5$. The maximum Doppler shift for generating time-varying TDL channels is determined based on the velocity $v\in \{100, 300\}$ km/h. The transmission frame shown in Fig. \ref{fig_pilot_pattern}(b) is adopted, where ${N_\text{p}}$ out of 14 OFDM symbols are allocated for pilot transmission. Traditional LMMSE and EP methods are employed in the first layer, consistent with the previous evaluations. For BEM-based receivers,\footnote{\emph{BEM-WLMMSE} integrates DPS and B-Spline basis functions to model fading across both time and frequency domains, utilizing time-frequency concatenation for data-aided computations. Conversely, the \emph{DPS-WLMMSE} method applies DPS basis functions to model frequency-domain autocorrelation at each time instant, offering a significant reduction in computational complexity.} the JCD loop requires $I = 11$ iterations,\footnote{Note that $I=11$ is selected for BEM-based methods to achieve optimal performance. However, performance at $I=2$ is also included in Fig. \ref{fig_comp_tradeoff} to ensure a fair comparison of computational complexity.} A total of $M_f = 9$ DPS sequences are chosen based on delay spread characteristics, while $M_t = 4$ is empirically selected to optimize the tradeoff between computational complexity and accuracy \cite{daCOMPKha}.

\subsubsection{Comparisons at Medium Speed}
\par
\begin{figure}[t]
  \begin{minipage}{3.5in}
    \centerline{\includegraphics[width=3.5in]{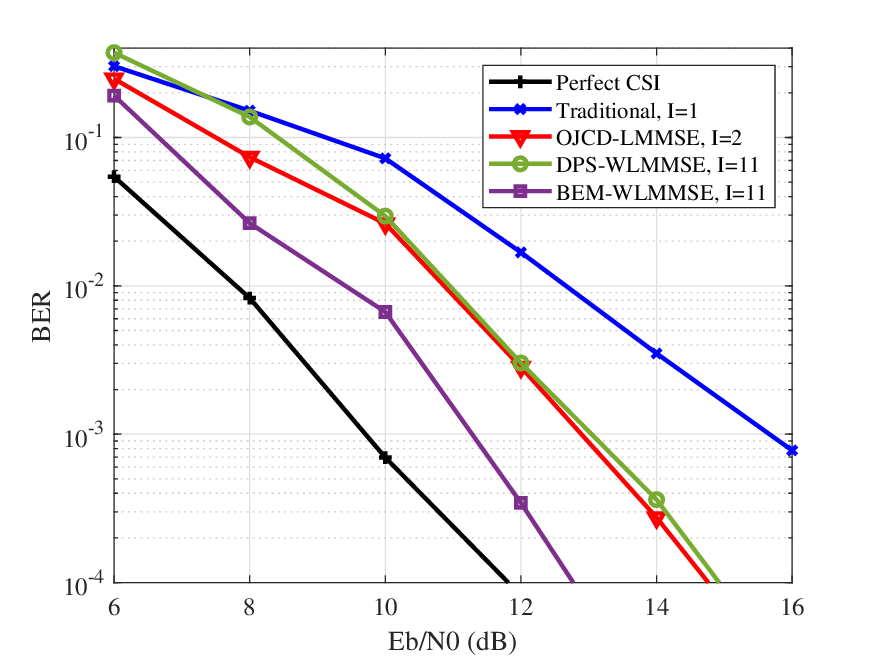}}
    \centerline{(a) $v=100$ km/h, $P=4$, $N_{\text{p}}=4$.}
  \end{minipage}
  \hfill
  \begin{minipage}{3.5in}
    \centerline{\includegraphics[width=3.5in]{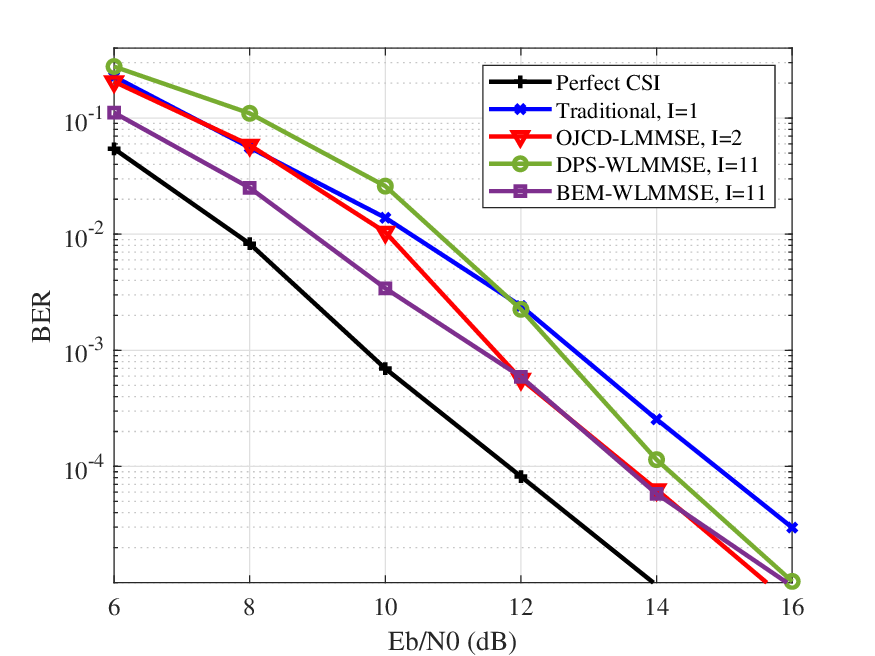}}
    \centerline{(b) $v=300$ km/h, $P=8$, $N_{\text{p}}=14$.}
  \end{minipage}
  \caption{~Detection performance comparison between the proposed receiver and BEM-based receivers under different time-varying scenarios in a $4 \times 4$ MIMO system with $K=128$ and $\rho=0.5$.}
  \vspace{-2em}
  \label{fig_comp_ber}
\end{figure}

The time-varying scenario with a velocity of $v=100$ km/h is considered first. A pilot size of $P = 4$ is selected to balance performance and transmission efficiency, while the remaining 112 subcarriers are reserved for data transmission. The detection accuracy in a coded system with QPSK modulation is presented in Fig. \ref{fig_comp_ber}(a). In both \emph{OJCD-LMMSE} and \emph{DPS-WLMMSE}-based receivers, data-aided estimates are computed at ${N_\text{p}} = 4$ symbols separately and subsequently interpolated in the time domain using cubic spline interpolation. As shown in Fig. \ref{fig_comp_ber}(a), the proposed \emph{OJCD-LMMSE}-based receiver outperforms the traditional method by more than 2.5 dB at a BER of $10^{-3}$. Additionally, the proposed receiver exhibits a performance advantage over the \emph{DPS-WLMMSE}-based receiver, which is attributed to the decision principle leading to incomplete utilization of data error information in the weighting matrix.
\par
In comparison to the \emph{BEM-WLMMSE}-based receiver, which fully exploits time-frequency correlation, \emph{OJCD-LMMSE} shows a performance gap of approximately 1.5 dB due to its use of an interpolator. While the \emph{BEM-WLMMSE} method provides superior data-aided improvement, it incurs considerable computational complexity due to the concatenation of dimensions. As illustrated in Fig. \ref{fig_comp_tradeoff}, the \emph{BEM-WLMMSE} method's superior BER performance is accompanied by a four-order magnitude increase in computational overhead. At $I = 2$, the performance gap narrows to approximately 0.75 dB, but the FLOP count remains three orders of magnitude higher. In contrast, \emph{OJCD-LMMSE} achieves a desirable balance between performance and complexity.

\subsubsection{Comparisons at High Speed}
\par
To further demonstrate the applicability of the proposed \emph{OJCD-LMMSE}-based receiver in high-mobility scenarios, it is evaluated under a time-varying condition with $v = 300$ km/h. With $P = 8$ and ${N_\text{p}} = 14$, the detection accuracy at the decoder output is presented in Fig. \ref{fig_comp_ber}(b). The results show that \emph{OJCD-LMMSE} achieves a gain of approximately 1.3 dB over the traditional method at a target BER of $10^{-4}$ and outperforms the \emph{DPS-WLMMSE}-based receiver, validating the effectiveness of incorporating detection error statistics.
\par
\begin{figure}[t]
  \begin{minipage}{3.5in}
    \centerline{\includegraphics[width=3.5in]{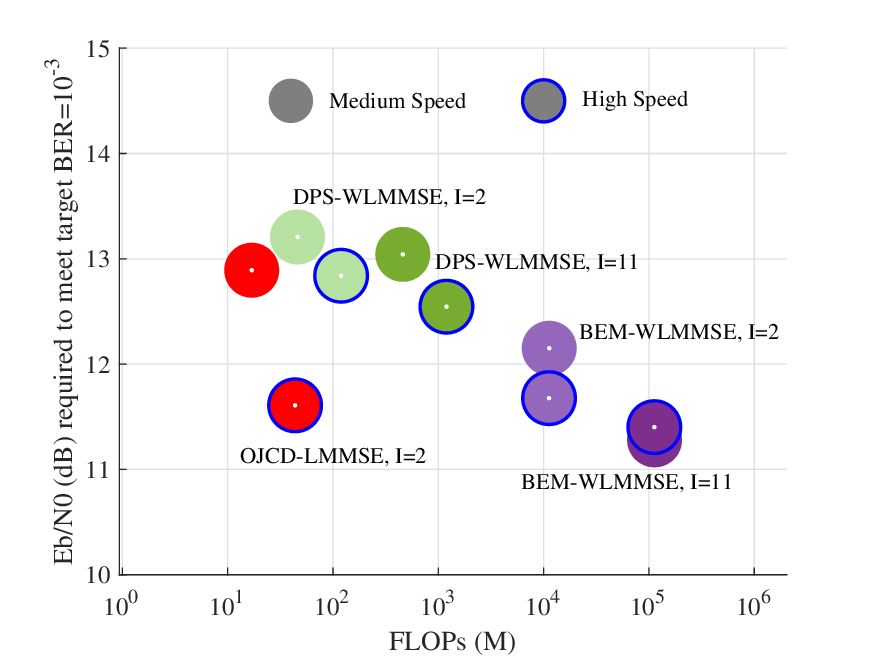}}
  \end{minipage}
  \caption{~Performance-complexity trade-off for data-aided receivers evaluated under a $4 \times 4$ MIMO system with $K=128$ and $\rho=0.5$. For the medium speed scenario ($v=100$ km/h), $P=4$ and $N_{\text{p}}=4$ are set. For the high speed scenario ($v=300$ km/h), $P=8$ and $N_{\text{p}}=14$ are set.}
  \vspace{-1em}
  \label{fig_comp_tradeoff}
\end{figure}

Notably, the \emph{BEM-WLMMSE} method shows reduced performance at higher SNRs due to an inadequate $M_t = 4$ setting for high mobility. Nonetheless, the proposed \emph{OJCD-LMMSE}-based receiver remains robust, performing computations for each OFDM symbol. As shown in Fig. \ref{fig_comp_ber}(b), the performance gap between \emph{OJCD-LMMSE} and \emph{BEM-WLMMSE} narrows to 0.2 dB at a BER of $10^{-3}$. Fig. \ref{fig_comp_tradeoff} further highlights the superior computational efficiency of \emph{OJCD-LMMSE}.

\vspace{-1.0em}
\section{Conclusion}
\label{sec_conclusion}
\par
We derived the general form of the data-aided LMMSE channel estimation within a JCD structure. This approach, utilizing detected symbols output by the EP detector for refined channel estimates, constitutes the \emph{MJCD-LMMSE} based structure and demonstrates remarkable performance improvement compared to traditional designs. A low-complexity equivalent algorithm, \emph{OJCD-LMMSE}, has been proposed as a substitute, effectively reducing complexity without sacrificing performance. Simulation results under various conditions, including different MIMO-OFDM configurations, pilot information, and time-varying characteristics, validate the accuracy advancement that the \emph{OJCD-LMMSE}-based receiver offers over traditional designs. Furthermore, our proposed \emph{OJCD-LMMSE} based receiver exhibits superiority over BEM-based data-aided receivers in balancing computational overhead and detection accuracy.

\appendices
\vspace{-1.0em}
\section{Derivation of MJCD-LMMSE}
\par
Given the assumptions on statistical properties of all concerned random variables listed in Section \ref{sec_MIMO_OFDM_CE}-A, (a)-(c), the deduction can be operated as follows. The derivation of ${{\mathbf{C}}_{{\mathbf{yh}}}}$ is firstly operated: 
  \begin{align}
    {{\mathbf{C}}_{{\mathbf{yh}}}} &= \mathbb{E}_{h,x,w} \left\{ {{\mathbf{y}}{{\mathbf{h}}^{\mathrm{H}}}} \right\} \notag \\
    &= \mathbb{E}_x \left\{ {\mathbf{X}} \right\} \mathbb{E}_h \left\{ {{\mathbf{h}}{{\mathbf{h}}^{\mathrm{H}}}} \right\} + \mathbb{E}_w \left\{ {\mathbf{w}} \right\} \mathbb{E}_h \left\{ {{{\mathbf{h}}^{\mathrm{H}}}} \right\} \notag \\
    &= {\hat{\mathbf{X}}}{{\mathbf{C}}_{{\mathbf{hh}}}}. \label{deriv_Cyh}
  \end{align} 
Note that ${\hat{\mathbf{X}}}$ is used to represent $\mathbb{E} \{ {\mathbf{X}} \}$ since $\mathbb{E}\{ {{x_i}} \} = {\hat x_i}$. Similarly, ${\hat{\mathbf{x}}}=\mathbb{E} \{ {\mathbf{x}} \}$. The derivation of ${{\mathbf{C}}_{{\mathbf{yy}}}}$ 
  \begin{align}
    {{\mathbf{C}}_{{\mathbf{yy}}}} &= \mathbb{E}_{h,x,w} \left\{ {{\mathbf{y}}{{\mathbf{y}}^{\mathrm{H}}}} \right\} \notag \\ 
    &= \mathbb{E}_{h,x} \left\{ {{\mathbf{Xh}}{{\mathbf{h}}^{\text{H}}}{{\mathbf{X}}^{\text{H}}}} \right\} + \mathbb{E}_x \left\{ {\mathbf{X}} \right\} \mathbb{E}_h \left\{ {\mathbf{h}} \right\} \mathbb{E}_w \left\{ {{{\mathbf{w}}^{\text{H}}}} \right\} \notag \\
    &~~ + \mathbb{E}_w \left\{ {\mathbf{w}} \right\} \mathbb{E}_h \left\{ {{{\mathbf{h}}^{\text{H}}}} \right\}\mathbb{E}_x \left\{ {{{\mathbf{X}}^{\text{H}}}} \right\} + \mathbb{E}_w \left\{ {{\mathbf{w}}{{\mathbf{w}}^{\text{H}}}} \right\} \notag \\
    &= \mathbb{E}_x \left\{ {{\mathbf{X}}{{\mathbf{C}}_{{\mathbf{hh}}}}{{\mathbf{X}}^{\mathrm{H}}}} \right\} + \sigma _w^2{\mathbf{I}}, \label{deriv_Cyy_step1}
  \end{align} 
\begin{figure*}[ht]
  \centering
  \begin{equation}
    \begin{aligned}
      \mathbb{E}_x \left\{ {{\mathbf{X}}{{\mathbf{C}}_{{\mathbf{hh}}}}{{\mathbf{X}}^{\text{H}}}} \right\} &= \mathbb{E}_x \left\{ {\left( {{{\mathbf{I}}_{{N_{\text{R}}}}} \otimes \left( {\left( {{{\mathbf{1}}_{1 \times {N_{\text{T}}}}} \otimes {{\mathbf{I}}_{\left( {K - P} \right)}}} \right) \cdot {\text{diag}}\left( {\mathbf{x}} \right)} \right)} \right){{\mathbf{C}}_{{\mathbf{hh}}}}{{\left( {{{\mathbf{I}}_{{N_{\text{R}}}}} \otimes \left( {\left( {{{\mathbf{1}}_{1 \times {N_{\text{T}}}}} \otimes {{\mathbf{I}}_{\left( {K - P} \right)}}} \right) \cdot {\text{diag}}\left( {\mathbf{x}} \right)} \right)} \right)}^{\text{H}}}} \right\} \\
      &= \mathbb{E}_x \left\{ {\left( {\left( {{{\mathbf{I}}_{{N_{\text{R}}}}} \otimes {{\mathbf{1}}_{1 \times {N_{\text{T}}}}} \otimes {{\mathbf{I}}_{\left( {K - P} \right)}}} \right) \cdot \left( {{{\mathbf{I}}_{{N_{\text{R}}}}} \otimes {\text{diag}}\left( {\mathbf{x}} \right)} \right)} \right){{\mathbf{C}}_{{\mathbf{hh}}}}{{\left( {\left( {{{\mathbf{I}}_{{N_{\text{R}}}}} \otimes {{\mathbf{1}}_{1 \times {N_{\text{T}}}}} \otimes {{\mathbf{I}}_{\left( {K - P} \right)}}} \right) \cdot \left( {{{\mathbf{I}}_{{N_{\text{R}}}}} \otimes {\text{diag}}\left( {\mathbf{x}} \right)} \right)} \right)}^{\text{H}}}} \right\} \\
      &= \left( {{{\mathbf{I}}_{{N_{\text{R}}}}} \otimes {{\mathbf{1}}_{1 \times {N_{\text{T}}}}} \otimes {{\mathbf{I}}_{\left( {K - P} \right)}}} \right) \mathbb{E}_x \left\{ {\left( {{{\mathbf{I}}_{{N_{\text{R}}}}} \otimes {\text{diag}}\left( {\mathbf{x}} \right)} \right){{\mathbf{C}}_{{\mathbf{hh}}}}{{\left( {{{\mathbf{I}}_{{N_{\text{R}}}}} \otimes {\text{diag}}\left( {\mathbf{x}} \right)} \right)}^{\text{H}}}} \right\}{\left( {{{\mathbf{I}}_{{N_{\text{R}}}}} \otimes {{\mathbf{1}}_{1 \times {N_{\text{T}}}}} \otimes {{\mathbf{I}}_{\left( {K - P} \right)}}} \right)^{\text{H}}} \\
    \end{aligned}
    \label{deriv_Cyy_step2}
  \end{equation}
  \vspace*{1ex}
  \begin{equation}
    \mathbb{E}_x \left\{ {{\mathbf{X}}{{\mathbf{C}}_{{\mathbf{hh}}}}{{\mathbf{X}}^{\text{H}}}} \right\} = {\hat{\mathbf{X}}}{{\mathbf{C}}_{{\mathbf{hh}}}}{{\hat{\mathbf{X}}}^{\text{H}}} + \left( {{{\mathbf{I}}_{{N_{\text{R}}}}} \otimes {{\mathbf{1}}_{1 \times {N_{\text{T}}}}} \otimes {{\mathbf{I}}_{\left( {K - P} \right)}}} \right)\left( {{{\mathbf{C}}_{{\mathbf{hh}}}} \odot \left( {{{\mathbf{1}}_{{N_{\text{R}}} \times {N_{\text{R}}}}} \otimes {\text{diag}}\left( {\mathbf{v}} \right)} \right)} \right){\left( {{{\mathbf{I}}_{{N_{\text{R}}}}} \otimes {{\mathbf{1}}_{1 \times {N_{\text{T}}}}} \otimes {{\mathbf{I}}_{\left( {K - P} \right)}}} \right)^{\text{H}}}
    \label{deriv_Cyy_step4}
  \end{equation}
  \vspace*{-1em}
  \hrulefill
\end{figure*}
\newline where $\mathbb{E}_x \{ {{\mathbf{X}}{{\mathbf{C}}_{{\mathbf{hh}}}}{{\mathbf{X}}^{\mathrm{H}}}} \}$ in (\ref{deriv_Cyy_step1}) is derived using the property of Kronecker product firstly, as shown in (\ref{deriv_Cyy_step2}). The expectation term in (\ref{deriv_Cyy_step2}) is equivalently represented using the partitioning of matrix as
  \begin{gather}
    \left( {{{\mathbf{I}}_{{N_{\text{R}}}}} \otimes {\text{diag}}\left( {\mathbf{x}} \right)} \right){{\mathbf{C}}_{{\mathbf{hh}}}}{\left( {{{\mathbf{I}}_{{N_{\text{R}}}}} \otimes {\text{diag}}\left( {\mathbf{x}} \right)} \right)^{\mathrm{H}}} \hfill \notag \\
     = {\left[ {{\text{diag}}\left( {\mathbf{x}} \right){{\mathbf{C}}_{ij}}{\text{diag}}{{\left( {\mathbf{x}} \right)}^{\mathrm{H}}}} \right]_{{N_{\text{R}}} \times {N_{\text{R}}}}}, \hfill  
  \end{gather} 
\newline where ${{\mathbf{C}}_{ij}} \in {\mathbb{C}^{{N_{\text{T}}}( {K - P} ) \times {N_{\text{T}}}( {K - P} )}}$ denotes the corresponding element after partitioning ${{\mathbf{C}}_{{\mathbf{hh}}}}$ into $N_{\text{R}}^2$ blocks such that ${{\mathbf{C}}_{{\mathbf{hh}}}} = {[ {{{\mathbf{C}}_{ij}}} ]_{{N_{\text{R}}} \times {N_{\text{R}}}}}$. Due to the statistical property of the transmitted symbols in (c), the expectation of the partitioned element is
\begin{equation}
  \begin{gathered}
    \mathbb{E}_x \left\{ {{\text{diag}}\left( {\mathbf{x}} \right){{\mathbf{C}}_{ij}}{\text{diag}}{{\left( {\mathbf{x}} \right)}^{\mathrm{H}}}} \right\} \hfill \\
     = {\text{diag}}\left( {\hat{\mathbf{x}}} \right){{\mathbf{C}}_{ij}}{\text{diag}}{\left( {\hat{\mathbf{x}}} \right)^{\mathrm{H}}} + {{\mathbf{C}}_{ij}} \odot {\text{diag}}\left( {\mathbf{v}} \right), \hfill \\
  \end{gathered} 
\end{equation}
where ${\mathbf{v}} = [ {{v_1}, \ldots ,{v_{{N_{\text{T}}}( {K - P} )}}} ]$. Therefore, the expectation of the integral matrix in (\ref{deriv_Cyy_step2}) can be expressed as
\begin{equation}
  \begin{gathered}
    \mathbb{E}_x \left\{ {\left( {{{\mathbf{I}}_{{N_{\text{R}}}}} \otimes {\text{diag}}\left( {\mathbf{x}} \right)} \right){{\mathbf{C}}_{{\mathbf{hh}}}}{{\left( {{{\mathbf{I}}_{{N_{\text{R}}}}} \otimes {\text{diag}}\left( {\mathbf{x}} \right)} \right)}^{\mathrm{H}}}} \right\} \hfill \\
     = \left( {{{\mathbf{I}}_{{N_{\text{R}}}}} \otimes {\text{diag}}\left( {\hat{\mathbf{x}}} \right)} \right){{\mathbf{C}}_{{\mathbf{hh}}}}{\left( {{{\mathbf{I}}_{{N_{\text{R}}}}} \otimes {\text{diag}}\left( {\hat{\mathbf{x}}} \right)} \right)^{\mathrm{H}}} \hfill \\
     ~~+ {{\mathbf{C}}_{{\mathbf{hh}}}} \odot \left( {{{\mathbf{1}}_{{N_{\text{R}}} \times {N_{\text{R}}}}} \otimes {\text{diag}}\left( {\mathbf{v}} \right)} \right). \hfill \\
  \end{gathered} 
\end{equation}
After substituting the above expression into (\ref{deriv_Cyy_step2}), the expectation term to be deduced is concluded as (\ref{deriv_Cyy_step4}), and the complete equation is summarized in (\ref{equ_mimo_ofdm_jcd_lmmse}).

\vspace{-1.0em}
\section{Derivation of OJCD-LMMSE}
According to (\ref{equ_W_LMMSEnew}), ${{\mathbf{R}}_{{{\mathbf{h}}^{{\text{d}}}}{{\mathbf{h}}^{{\text{LS,new}}}}}}$ and ${\mathbf{R}}_{{{\mathbf{h}}^{{\text{LS,new}}}}{{\mathbf{h}}^{{\text{LS,new}}}}}$ are required and defined for the computation of ${\mathbf{W}}_{{\text{LMMSE}}}^{{\text{new}}}$. Note that the assumptions and statistical properties concluded in Section \ref{sec_MIMO_OFDM_CE} are still applicable. Besides, according to (\ref{equ_def_error_ofdm}), the estimation error of $\hat{\mathbf{h}}_{m,n}^{\text{d}}$ should be taken into consideration, and the estimated channel coefficients are calculated from traditional LMMSE algorithm adopted in JCD-1 as shown in (\ref{equ_h_mn_LS})-(\ref{equ_h_mn_LMMSE}), which can also be represented as
\begin{equation}
  \begin{aligned}
    {\mathbf{h}}_{m,n}^{{\text{LMMSE}}} &= {{\mathbf{W}}_{{\text{LMMSE}}}}{\mathbf{h}}_{m,n}^{{\text{LS}}} \\
    &= {{\mathbf{W}}_{{\text{LMMSE}}}}\left( {{\mathbf{h}}_{m,n}^{\text{p}} + {{\left( {{\mathbf{X}}_n^{\text{p}}} \right)}^{ - 1}}{\mathbf{w}}_{m,n}^{\text{p}}} \right). \nonumber
  \end{aligned}
\end{equation}
Therefore, the following relation exists:
\begin{equation}
  \hat{\mathbf{h}}_{m,n}^{\text{d}} = {\mathbf{W}}_{{\text{LMMSE}}}^{\text{d}}\left( {{\mathbf{h}}_{m,n}^{\text{p}} + {{\left( {{\mathbf{X}}_n^{\text{p}}} \right)}^{ - 1}}{\mathbf{w}}_{m,n}^{\text{p}}} \right),
  \label{deriv_def_hhatd}
\end{equation}
where ${\mathbf{W}}_{{\text{LMMSE}}}^{\text{d}}$ refers to extracting specific rows from ${{\mathbf{W}}_{{\text{LMMSE}}}}$ according to the index set for data subcarriers $\{ {{d_k}} \}_{k = 1}^{K-P}$. ${\mathbf{W}}_{{\text{LMMSE}}}^{\text{d}}$ and ${{\mathbf{X}}_n^{\text{p}}}$ are treated as known constants. Moreover, for simplicity, ${\mathbf{W}}_1 = {\mathbf{W}}_{{\text{LMMSE}}}^{\text{d}}$ is defined. The correlation properties between OFDM channels corresponding to different transmitting antennas, which can all be computed from ${{\mathbf{R}}={\mathbf{R}}_{\text{t}} \otimes {\mathbf{R}}^{\text{Freq}}}$, are defined as follows:
\begin{equation}
  \begin{aligned}
    {{\mathbf{R}}_{{{\mathbf{h}}^{\text{d}}}{{\mathbf{h}}^{\text{d}}}}}\left( {{n_1},{n_2}} \right) &= \mathbb{E}\left\{ {{\mathbf{h}}_{m,{n_1}}^{\text{d}}{{\left( {{\mathbf{h}}_{m,{n_2}}^{\text{d}}} \right)}^{\mathrm{H}}}} \right\}, \\
    {{\mathbf{R}}_{{{\mathbf{h}}^{\text{d}}}{{\mathbf{h}}^{\text{p}}}}}\left( {{n_1},{n_2}} \right) &= \mathbb{E}\left\{ {{\mathbf{h}}_{m,{n_1}}^{\text{d}}{{\left( {{\mathbf{h}}_{m,{n_2}}^{\text{p}}} \right)}^{\mathrm{H}}}} \right\}, \\
    {{\mathbf{R}}_{{{\mathbf{h}}^{\text{p}}}{{\mathbf{h}}^{\text{p}}}}}\left( {{n_1},{n_2}} \right) &= \mathbb{E}\left\{ {{\mathbf{h}}_{m,{n_1}}^{\text{p}}{{\left( {{\mathbf{h}}_{m,{n_2}}^{\text{p}}} \right)}^{\mathrm{H}}}} \right\},
  \end{aligned}
  \label{deriv_def_Rs}
\end{equation}
where ${n_1},{n_2} \in \{ {1, \ldots ,{N_{\text{T}}}} \}$. Note that if ${n_1}={n_2}$, ${{\mathbf{R}}_{{{\mathbf{h}}^{\text{d}}}{{\mathbf{h}}^{\text{d}}}}}( {{n_1},{n_2}} )$ can be simplified as ${{\mathbf{R}}_{{{\mathbf{h}}^{\text{d}}}{{\mathbf{h}}^{\text{d}}}}}$.
\par
The deduction is now operated. By plugging in (\ref{equ_h_mn_LSnew2}), the derivation of ${{\mathbf{R}}_{{{\mathbf{h}}^{{\text{d}}}}{{\mathbf{h}}^{{\text{LS,new}}}}}}$ is preliminarily operated as
\begin{equation}
  \begin{aligned}
    {{\mathbf{R}}_{{{\mathbf{h}}^{{\text{d}}}}{{\mathbf{h}}^{{\text{LS,new}}}}}}\left( n \right) &= \mathbb{E}_{h,x,w} \left\{ {{\mathbf{h}}_{m,n}^{\text{d}}{{\left( {{\mathbf{h}}_{m,n}^{{\text{LS,new}}}} \right)}^{\mathrm{H}}}} \right\} = {{\mathbf{R}}_{{{\mathbf{h}}^{\text{d}}}{{\mathbf{h}}^{\text{d}}}}} + {{\mathbf{B}}_n}, \\
    {{\mathbf{B}}_n} &\triangleq \mathbb{E}_{h,x,w} \left\{ {{\mathbf{h}}_{m,n}^{\text{d}}{\mathbf{Z}}_{m,n}^{\mathrm{H}}} \right\}{\left( {{{\left( {{\hat{\mathbf{X}}}_n^{\text{d}}} \right)}^{ - 1}}} \right)^{\mathrm{H}}}.
  \end{aligned}
  \label{deriv_Rhhls_step1}
\end{equation}

And ${\mathbf{R}}_{{{\mathbf{h}}^{{\text{LS,new}}}}{{\mathbf{h}}^{{\text{LS,new}}}}}$ is preliminarily deduced as
\begin{equation}
  \begin{aligned}
    {{\mathbf{R}}_{{{\mathbf{h}}^{{\text{LS,new}}}}{{\mathbf{h}}^{{\text{LS,new}}}}}}\left( n \right) &= \mathbb{E}_{h,x,w} \left\{ {\left( {{\mathbf{h}}_{m,n}^{{\text{LS,new}}}} \right){{\left( {{\mathbf{h}}_{m,n}^{{\text{LS,new}}}} \right)}^{\mathrm{H}}}} \right\} \\
    &= {{\mathbf{R}}_{{{\mathbf{h}}^{\text{d}}}{{\mathbf{h}}^{\text{d}}}}} + {{\mathbf{B}}_n} + {\mathbf{B}}_n^{\mathrm{H}} \\
    &~~ + {\left( {{\hat{\mathbf{X}}}_n^{\text{d}}} \right)^{ - 1}}{\bm{\Sigma} _n}{{{{\left( {{\hat{\mathbf{X}}}_n^{\text{d}}} \right)}^{ - 1}}} ^{\mathrm{H}}} \\
    {\bm{\Sigma} _n} &\triangleq \mathbb{E}_{h,x,w} \left\{ {{{\mathbf{Z}}_{m,n}}{\mathbf{Z}}_{m,n}^{\mathrm{H}}} \right\}.
  \end{aligned}
  \label{deriv_Rhlshls_step1}
\end{equation}

\par
Consequently, for subsequent inference in (\ref{deriv_Rhhls_step1}) and (\ref{deriv_Rhlshls_step1}), ${{\mathbf{B}}_n}$ and ${\bm{\Sigma} _n}$ must be deduced first. ${{\mathbf{B}}_n}$ is unfolded according to (\ref{equ_Z_mn}), and is further simplified using the zero-mean property of $w$ and $\Delta e$:
\begin{equation}
  \begin{aligned}
    {{\mathbf{B}}_n} &= \sum\limits_{\substack{n' = 1\\n' \ne n}}^{{N_{\text{T}}}} {\mathbb{E}\left\{ {{\mathbf{h}}_{m,n}^{\text{d}}{{\left( {{\mathbf{\Delta h}}_{m,n'}^{\text{d}}} \right)}^{\mathrm{H}}}} \right\}{{\left( {{\hat{\mathbf{X}}}_{n'}^{\text{d}}} \right)}^{\mathrm{H}}}} {{{{\left( {{\hat{\mathbf{X}}}_n^{\text{d}}} \right)}^{ - 1}}} ^{\mathrm{H}}} \\
    &= \sum\limits_{\substack{n' = 1\\n' \ne n}}^{{N_{\text{T}}}} {{{\mathbf{R}}_{{{\mathbf{h}}^{\text{d}}}{{\mathbf{h}}^{\text{d}}}}}\left( {n,n'} \right){{\left( {{\hat{\mathbf{X}}}_{n'}^{\text{d}}} \right)}^{\mathrm{H}}}} {\left( {{{\left( {{\hat{\mathbf{X}}}_n^{\text{d}}} \right)}^{ - 1}}} \right)^{\mathrm{H}}} \\
    &~~ - \sum\limits_{\substack{n' = 1\\n' \ne n}}^{{N_{\text{T}}}} {\mathbb{E}\left\{ {{\mathbf{h}}_{m,n}^{\text{d}}{{\left( {\hat{\mathbf{h}}_{m,n'}^{\text{d}}} \right)}^{\mathrm{H}}}} \right\}{{\left( {{\hat{\mathbf{X}}}_{n'}^{\text{d}}} \right)}^{\mathrm{H}}}} {\left( {{{\left( {{\hat{\mathbf{X}}}_n^{\text{d}}} \right)}^{ - 1}}} \right)^{\mathrm{H}}}.
  \end{aligned}
  \label{deriv_Rhhls_step2}
\end{equation}

The remaining expectation term in (\ref{deriv_Rhhls_step2}) is unfolded due to the representation in (\ref{deriv_def_hhatd}), and is consequently simplified using the zero-mean property of $w$:
\begin{equation}
  \mathbb{E}\left\{ {{\mathbf{h}}_{m,n}^{\text{d}}{{\left( {\hat{\mathbf{h}}_{m,n'}^{\text{d}}} \right)}^{\mathrm{H}}}} \right\} = {{\mathbf{R}}_{{{\mathbf{h}}^{\text{d}}}{{\mathbf{h}}^{\text{p}}}}}\left( {n,n'} \right){ {\mathbf{W}}_{1}^{\mathrm{H}}}. \nonumber
\end{equation}

${{\mathbf{B}}_n}$ can finally be represented as
\begin{equation}
  {{\mathbf{B}}_n} = \sum\limits_{\substack{n' = 1\\n' \ne n}}^{{N_{\text{T}}}} {\left( \begin{gathered}
    {{\mathbf{R}}_{{{\mathbf{h}}^{\text{d}}}{{\mathbf{h}}^{\text{d}}}}}\left( {n,n'} \right) \hfill \\
     - {{\mathbf{R}}_{{{\mathbf{h}}^{\text{d}}}{{\mathbf{h}}^{\text{p}}}}}\left( {n,n'} \right){{\mathbf{W}}_{1}^{\mathrm{H}}} \hfill \\
  \end{gathered}  \right){{\left( {{\hat{\mathbf{X}}}_{n'}^{\text{d}}} \right)}^{\mathrm{H}}}} {\left( {{{\left( {{\hat{\mathbf{X}}}_n^{\text{d}}} \right)}^{ - 1}}} \right)^{\mathrm{H}}}.
  \label{deriv_Rhhls_step3}
\end{equation}
\par
Likewise, ${\bm{\Sigma} _n}$ is deduced as follows. The defined expression is firstly unfolded using (\ref{equ_Z_mn}), and subsequently represented according to the second-order statistical property of $w$ and $\Delta e$:

\begin{equation}
  \begin{aligned}
    {\bm{\Sigma} _n} &= \sum\limits_{\substack{{{n}_1'} = 1\\{{n}_1'} \ne n}}^{{N_{\text{T}}}} {\sum\limits_{\substack{{{n}_2'} = 1\\{{n}_2'} \ne n}}^{{N_{\text{T}}}} {{\hat{\mathbf{X}}}_{{{n}_1'}}^{\text{d}}\mathbb{E}\left\{ {{\mathbf{\Delta h}}_{m,{{n}_1'}}^{\text{d}}{{\left( {{\mathbf{\Delta h}}_{m,{{n}_2'}}^{\text{d}}} \right)}^{\mathrm{H}}}} \right\}{{\left( {{\hat{\mathbf{X}}}_{{{n}_2'}}^{\text{d}}} \right)}^{\mathrm{H}}}} } \\
    &~~ + {{\mathbf{R}}_{{{\mathbf{h}}^{\text{d}}}{{\mathbf{h}}^{\text{d}}}}} \odot \sum\limits_{n = 1}^{{N_{\text{T}}}} {{{\mathbf{V}}_n}}  + \sigma _w^2{\mathbf{I}},
  \end{aligned}
  \label{deriv_Rhlshls_step2}
\end{equation}
and the expectation term in (\ref{deriv_Rhlshls_step2}) is again expanded using (\ref{deriv_def_hhatd})

\begin{equation}
  \begin{aligned}
    & \mathbb{E}\left\{ {{\mathbf{\Delta h}}_{m,{{n}_1'}}^{\text{d}}{{\left( {{\mathbf{\Delta h}}_{m,{{n}_2'}}^{\text{d}}} \right)}^{\mathrm{H}}}} \right\} \\
    &= {{\mathbf{R}}_{{{\mathbf{h}}^{\text{d}}}{{\mathbf{h}}^{\text{d}}}}}\left( {{{n}_1'},{{n}_2'}} \right) - {{\mathbf{R}}_{{{\mathbf{h}}^{\text{d}}}{{\mathbf{h}}^{\text{p}}}}}\left( {{{n}_1'},{{n}_2'}} \right){\mathbf{W}}_{\text{1}}^{\mathrm{H}} \\
    &~~ - {{\mathbf{W}}_1}{{\mathbf{R}}_{{{\mathbf{h}}^{\text{p}}}{{\mathbf{h}}^{\text{d}}}}}\left( {{{n}_1'},{{n}_2'}} \right) + {{\mathbf{W}}_1}{{\mathbf{R}}_{{{\mathbf{h}}^{\text{p}}}{{\mathbf{h}}^{\text{p}}}}}\left( {{{n}_1'},{{n}_2'}} \right){\mathbf{W}}_{\text{1}}^{\mathrm{H}} \\
    &~~ + \sigma _w^2{{\mathbf{W}}_1}{\left( {{\mathbf{X}}_{{{n}_1'}}^{\text{p}}} \right)^{ - 1}}{\left( {{{\left( {{\mathbf{X}}_{{{n}_1'}}^{\text{p}}} \right)}^{ - 1}}} \right)^{\mathrm{H}}}{\mathbf{W}}_{\text{1}}^{\mathrm{H}}. \\
  \end{aligned} \nonumber
\end{equation}
${\bm{\Sigma} _n}$ can be summarized as
\begin{equation}
  \begin{gathered}
    {\bm{\Sigma} _n} = \sum\limits_{\substack{{{n}_1'} = 1\\{{n}_1'} \ne n}}^{{N_{\text{T}}}} {\sum\limits_{\substack{{{n}_2'} = 1\\{{n}_2'} \ne n}}^{{N_{\text{T}}}} {{\left(
      \begin{gathered}
      {{\mathbf{R}}_{{{\mathbf{h}}^{\text{d}}}{{\mathbf{h}}^{\text{d}}}}}\left( {{{n}_1'},{{n}_2'}} \right) \hfill \\
       - {{\mathbf{V}}^{\text{A}}}\left( {{{n}_1'},{{n}_2'}} \right) \hfill \\
       + {{\mathbf{V}}^{\text{B}}}\left( {{{n}_1'},{{n}_2'}} \right) \hfill \\
      \end{gathered}  \right)} \odot {{\mathbf{V}}^{\text{x}}}\left( {{{n}_1'},{{n}_2'}} \right)} }  \\
     ~~~~~~+ \sigma _w^2\sum\limits_{\substack{{{n'}} = 1\\{{n'}} \ne n}}^{{N_{\text{T}}}} {{{\mathbf{V}}^{\text{C}}}\left( {{{n'}}} \right) \odot {{\mathbf{V}}^{\text{x}}}\left( {{{n'}},{{n'}}} \right)}  + {{\mathbf{V}}_{\text{D}}} + \sigma _w^2{\mathbf{I}}, \\
  \end{gathered}
  \label{deriv_Rhlshls_step3}
\end{equation}
where
\begin{equation}
  \begin{aligned}
    {{\mathbf{V}}^{\text{A}}}\left( {{{n}_1'},{{n}_2'}} \right) &\triangleq {{\mathbf{R}}_{{{\mathbf{h}}^{\text{d}}}{{\mathbf{h}}^{\text{p}}}}}\left( {{{n}_1'},{{n}_2'}} \right){\mathbf{W}}_{\text{1}}^{\mathrm{H}} + {{\mathbf{W}}_1}{\mathbf{R}}_{{{\mathbf{h}}^{\text{d}}}{{\mathbf{h}}^{\text{p}}}}^{\mathrm{H}}\left( {{{n}_1'},{{n}_2'}} \right), \\
    {{\mathbf{V}}^{\text{B}}}\left( {{{n}_1'},{{n}_2'}} \right) &\triangleq {{\mathbf{W}}_1}{{\mathbf{R}}_{{{\mathbf{h}}^{\text{p}}}{{\mathbf{h}}^{\text{p}}}}}\left( {{{n}_1'},{{n}_2'}} \right){\mathbf{W}}_{\text{1}}^{\mathrm{H}}, \\
    {{\mathbf{V}}^{\text{x}}}\left( {{{n}_1'},{{n}_2'}} \right) &\triangleq {\hat{\mathbf{x}}}_{{{n}_1'}}^{\text{d}}{\left( {{\hat{\mathbf{x}}}_{{{n}_2'}}^{\text{d}}} \right)^{\mathrm{H}}}, \\
    {{\mathbf{V}}^{\text{C}}}\left( {{{n'}}} \right) &\triangleq \left( {{{\mathbf{W}}_1}{{\left( {{\mathbf{X}}_{{{n'}}}^{\text{p}}} \right)}^{ - 1}}} \right){\left( {{{\mathbf{W}}_1}{{\left( {{\mathbf{X}}_{{{n'}}}^{\text{p}}} \right)}^{ - 1}}} \right)^{\mathrm{H}}}, \\
    {{\mathbf{V}}_{\text{D}}} &\triangleq {{\mathbf{R}}_{{{\mathbf{h}}^{\text{d}}}{{\mathbf{h}}^{\text{d}}}}} \odot \sum\limits_{n = 1}^{{N_{\text{T}}}} {{{\mathbf{V}}_n}}. \nonumber
  \end{aligned}
\end{equation}

\bibliographystyle{IEEEtran}
\bibliography{referencePaper}
\end{document}